\documentclass[12pt]{article}

\usepackage{axodraw}
\usepackage{color}

\usepackage{amsmath}
\input{epsf}
\def\slash#1{\ooalign{$\hfil/\hfil$\crcr$#1$}} 

\def\td{{\widetilde \delta}}
\def\qt{{\bf q}_{\perp}}
\def\kt{{\bf k}_{\perp}}

\def\res{{\rm Res}_{ \{{\rm Im}\; q_0 < 0 \}}}
\def\reslc{{\rm Res}_{ \{{\rm Im}\; q_+ < 0 \}}}

\def\b{{\rm Bub}}

\def\sg{{\rm sign}}

\def\bar{\overline}
\def\tr{{\rm Tr}}

\def\ep{\epsilon}

\def\beq{\begin{equation}}
\def\eeq{\end{equation}}
\def\beeq{\begin{eqnarray}}
\def\eeeq{\end{eqnarray}}

\def\to{\rightarrow}

\def\nn{\nonumber}

\def\ID{1 \kern -.45 em 1}

\begin{document}

\begin{titlepage}
\renewcommand{\thefootnote}{\fnsymbol{footnote}}
\begin{flushright}
     IFIC/08-21, IPPP/08/22, FERMILAB-PUB-08-092-T, SLAC-PUB-13218 \\
     arXiv:0804.3170 [hep-ph]
     \end{flushright}
\begin{center}
{\LARGE \bf
From loops to trees \\[1ex]
by-passing Feynman's theorem
}
\end{center}
\par \vspace{2mm}
\begin{center}
{\bf Stefano Catani~$^{(a)}$\footnote{E-mail: stefano.catani@fi.infn.it}, 
Tanju Gleisberg~$^{(b)}$\footnote{E-mail: tanju@slac.stanford.edu},
Frank Krauss~$^{(c)}$\footnote{E-mail: frank.krauss@durham.ac.uk}, \\ [1ex]
Germ\'an Rodrigo~$^{(d)}$\footnote{E-mail: german.rodrigo@ific.uv.es} and
Jan-Christopher Winter~$^{(e)}$\footnote{E-mail: jwinter@fnal.gov}
}
\vspace{5mm}

${}^{(a)}$INFN, Sezione di Firenze and
Dipartimento di Fisica, Universit\`a di Firenze,\\
I-50019 Sesto Fiorentino,
Florence, Italy \\
\vspace*{2mm}
${}^{(b)}$ Stanford Linear Accelerator Center, Stanford University \\
Stanford, CA 94309, USA \\
\vspace*{2mm}
${}^{(c)}$ Institute for Particle Physics Phenomenology, 
Durham University, \\ Durham DH1 3LE, UK \\
\vspace*{2mm}
${}^{(d)}$Instituto de F\'{\i}sica Corpuscular, 
CSIC-Universitat de Val\`encia, \\
Apartado de Correos 22085, E-46071 Valencia, Spain \\
\vspace*{2mm}
${}^{(e)}$Fermi National Accelerator Laboratory, Batavia, IL 60510, USA \\
\vspace*{2mm}

\end{center}

\par \vspace{2mm}
\begin{center} {\large \bf Abstract} \end{center}
\begin{quote}
\pretolerance 10000
We derive a duality relation between one-loop integrals and phase-space 
integrals emerging from them through single cuts. The duality relation 
is realized by a modification of the customary $+i0$ prescription of the 
Feynman propagators. The new prescription regularizing the propagators,
which we write in a Lorentz covariant form, compensates for the absence 
of multiple-cut contributions that appear in the Feynman Tree Theorem.
The duality relation can be applied to generic one-loop quantities in
any relativistic, local and unitary field theories.
We discuss in detail the duality that relates one-loop and tree-level 
Green's functions. We comment on applications
to the analytical calculation of
one-loop scattering amplitudes, and to the numerical evaluation of
cross-sections at next-to-leading order.
\end{quote}

\par \vspace{5mm}

\vspace*{\fill}
\begin{flushleft}
     arXiv:0804.3170 [hep-ph] \\ 20 April 2008 
\end{flushleft}
\end{titlepage}

\setcounter{footnote}{0}
\renewcommand{\thefootnote}{\fnsymbol{footnote}}

\section{Introduction}
\label{sec:intro}

The Feynman Tree Theorem (FTT) \cite{Feynman:1963ax,F2} applies to any 
(local and unitary) quantum field theories in Minkowsky space with an 
arbitrary number $d$ of space-time dimensions. It relates perturbative 
scattering amplitudes and Green's functions at the loop level with 
analogous quantities at the tree level. This relation follows from a basic 
and more elementary relation between loop integrals and phase-space integrals. 
Using this basic relation loop Feynman diagrams can be rewritten in terms 
of phase-space integrals of tree-level Feynman diagrams. The corresponding
tree-level Feynman diagrams are then obtained by considering {\em multiple} 
cuts (single cuts, double cuts, triple cuts and so forth) of the original 
loop Feynman diagram.

We have recently proposed a method \cite{meth,tanju,inprep} to 
numerically compute multi-leg one-loop cross sections in perturbative field 
theories. The starting point of this method is a {\em duality} relation 
between one-loop integrals and phase-space integrals. Although the analogy with
the FTT is quite close, there are important differences. The key difference 
is that the duality relation involves only {\em single} cuts of the 
one-loop Feynman diagrams.
Both the FTT and the duality relation can be derived by using the 
residue theorem\footnote{
	Within the context of loop integrals, the use of the residue 
	theorem has been considered many times in textbooks and in the literature.
}. 

In this paper, we illustrate and derive the duality relation.
Since the FTT has recently attracted a renewed interest 
\cite{Brandhuber:2005kd} in the context of twistor-inspired methods
\cite{Witten:2003nn,Cachazo:2004zb} to evaluate one-loop scattering 
amplitudes~\cite{Bern:2007dw}, we also
discuss its correspondence (including similarities and differences)
with the duality relation.

The outline of the paper is as follows. In Section~\ref{sec:not}, we introduce 
our notation. In Section~\ref{sec:ft}, we briefly recall how the FTT relates 
one-loop integrals with multiple-cut phase-space integrals. In 
Section~\ref{sec:dt}, we present one of the main results of this publication: 
we derive and illustrate the duality relation between one-loop integrals 
and single-cut phase-space integrals. We also prove that the duality relation 
requires to properly regularize propagators by a complex Lorentz-covariant
prescription, which is different from the customary $+i0$ prescription of the 
Feynman propagators. The duality is illustrated in Section~\ref{sec:2p} by
considering the two-point function as the simplest example application. 
The correspondence between the FTT and the duality relation is formalized in 
Section~\ref{sec:rel}. In Section~\ref{sec:dual}, 
we explore the one-to-one correspondence between one-loop Feynman integrals 
and single-cut integrals on more mathematical grounds, 
and establish a generalized duality relation. 
The treatment of particle masses (including complex masses of unstable
particles) when cutting loop integrals is discussed in
Section~\ref{sec:mass}. 
In Section~\ref{sec:gauge}, we analyze the effect of the gauge poles introduced 
by the propagators of the gauge fields in local gauge theories. In 
Section~\ref{sec:dam}, we discuss the extension of the duality relation 
to one-loop Green's 
functions and scattering amplitudes. 
Some final remarks are presented in Section~\ref{sec:fin}.
Details about the derivation of the duality relation by using the residue
theorem are discussed in Appendix~\ref{app:a}.
The proof of an algebraic relation is presented in Appendix~\ref{app:b}. 
Issues related to tadpole singularities are discussed in Appendix~\ref{app:c}.

\section{Notation}
\label{sec:not}

The FTT and the duality relation can be illustrated with no loss of generality
by considering their application to the basic ingredient of any one-loop
Feynman diagrams, namely a generic one-loop scalar integral $L^{(N)}$ with 
$N$ ($N \geq 2$) external legs. 

\begin{figure}[htb]
\begin{center}
\vspace*{8mm}
\begin{picture}(120,110)(0,-10)

\color{blue}
\SetWidth{1.2}
\BCirc(50,50){30}
\ArrowArc(50,50)(30,110,190)
\ArrowArc(50,50)(30,190,-50)
\ArrowArc(50,50)(30,-50,30)
\ArrowArc(50,50)(30,30,110)
\ArrowArc(50,50)(20,60,190)
\ArrowLine(39.74,78.19)(29.48,106.38)
\ArrowLine(75.98,65)(101.96,80)
\ArrowLine(69.28,27.01)(88.56,4.03)
\ArrowLine(20.45,44.79)(-9.09,39.58)
\Vertex(21.07,15.53){1.4}
\Vertex(34.60,7.71){1.4}
\Vertex(50,5){1.4}
\Text(44,55)[]{$q$}
\Text(40,110)[]{$p_1$}
\Text(65,90)[]{$q_1$}
\Text(115,80)[]{$p_2$}
\Text(93,45)[]{$q_2$}
\Text(10,65)[]{$q_N$}
\Text(0,30)[]{$p_N$}
\Text(100,0)[]{$p_3$}

\end{picture}
\end{center}
\vspace*{-6mm}
\caption{\label{f1loop} 
{\em Momentum configuration of the one-loop $N$-point scalar integral.
}}
\end{figure}
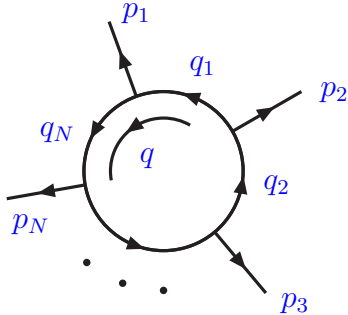

The momenta of the external legs are denoted by $p_1^\mu, p_2^\mu, \dots, p_N^\mu$
and are clockwise ordered (Fig.~\ref{f1loop}). All are taken as outgoing. To 
simplify the notation and the presentation, we also limit ourselves in the
beginning to considering massless internal lines only. Thus, the one-loop integral 
$L^{(N)}$ can in general be expressed as:  
\beq
\label{Ln}
L^{(N)}(p_1, p_2, \dots, p_N) = - i \, 
\int \frac{d^d q}{(2\pi)^d} \;
\prod_{i=1}^{N} \, \frac{1}{q_i^2+i 0} \;\;,
\eeq
where $q^\mu$ is the loop momentum (which flows anti-clockwise). The momenta
of the internal lines are denoted by $q_i^\mu$; they are given by
\beq
\label{defqi}
q_i = q + \sum_{k=1}^i p_k \;\;,
\eeq
and momentum conservation results in the constraint
\beq
\sum_{i=1}^N p_i = 0 \;\;.
\eeq
The value of the label $i$ of the external momenta is defined modulo 
$N$, i.e. $p_{N+i} \equiv p_{i}$.

The number of space-time dimensions is denoted by $d$ (the convention for the
Lorentz-indices adopted here is $\mu=0, 1, \dots, d-1$) with metric tensor 
$g^{\mu \nu} = {\rm diag}(+1,-1,\dots,-1)$. 
The space-time coordinates of any momentum $k_\mu$ are 
denoted as $k_\mu=(k_0, {\bf k})$, where $k_0$ is the energy (time component) of 
$k_\mu$. It is also convenient to introduce light-cone coordinates
$k_\mu=(k_+, \kt, k_-)$, where $k_{\pm} = (k_0 \pm k_{d-1})/{\sqrt 2}$.
Throughout the paper we consider loop integrals and phase-space integrals. If
the integrals are ultraviolet or infrared divergent, we always assume that they
are regularized by using analytic continuation in the number of space-time
dimensions (dimensional regularization). Therefore, $d$ is not fixed and 
does not necessarily have integer value.

We introduce 
the following shorthand notation:
\beq
- i \, \int \frac{d^d q}{(2\pi)^d} \;\;\cdots \equiv
\int_q \;\; \cdots \;\;.
\eeq
When we factorize off in a loop integral the integration over the momentum coordinate 
$q_0$ or $q_+$, we write
\beq
- i \, \int_{-\infty}^{+\infty} dq_0 \;\int \frac{d^{d-1}{\bf q}}{(2\pi)^d}
 \;\;\cdots \equiv \int dq_0 \;
\int_{\bf q} \;\; \cdots \;\;,
\eeq
and
\beq
\label{pslc}
- i \, \int_{-\infty}^{+\infty} dq_+ \;\int_{-\infty}^{+\infty} dq_-
\int \frac{d^{d-2}{\qt}}{(2\pi)^d}
 \;\;\cdots \equiv \int dq_+ \;
\int_{(q_-,\qt)} \;\; \cdots \;\;,
\eeq
respectively. The customary phase-space integral of a physical massless
particle with momentum $q$ (i.e. an on-shell particle with positive-definite 
energy: $q^2=0$, $q_0\geq 0$) reads
\beq
\label{psm}
\int \frac{d^d q}{(2\pi)^{d-1}} \;\theta(q_0) \;\delta(q^2) \;\;\cdots \equiv
\int_q \td(q) \;\;\cdots \;\;,
\eeq
where we have defined
\beq
\td(q) \equiv 2 \pi \, i \,\theta(q_0) \;\delta(q^2) 
= 2 \pi \, i \;\delta_+(q^2) \;\;.
\eeq

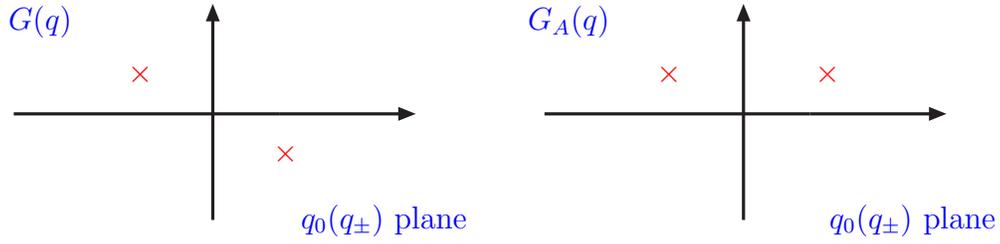
\begin{figure}[htb]
\begin{center}
\begin{picture}(350,110)(0,0)
\color{blue}
\SetWidth{1.2}
\Line(0,50)(100,50)
\Line(200,50)(300,50)
\LongArrow(100,50)(150,50)
\LongArrow(300,50)(350,50)
\LongArrow(75,10)(75,90)
\LongArrow(275,10)(275,90)
\Text(10,85)[]{$G(q)$}
\Text(210,85)[]{$G_A(q)$}
\Text(140,10)[]{$q_0 (q_{\pm})$ plane}
\Text(340,10)[]{$q_0 (q_{\pm})$ plane}
\color{red}
\Text(95,35)[]{$\times$}
\Text(40,65)[]{$\times$}
\Text(300,65)[]{$\times$}
\Text(240,65)[]{$\times$}
\end{picture}
\end{center}
\caption{\label{fvsa}
{\em Location of the particle poles of the Feynman 
(left)
and advanced 
(right) propagators, $G(q)$ and $G_A(q)$,
in the complex plane of the variable 
$q_0$ or $q_{\pm}$.
}}
\end{figure}

Using this shorthand notation, the one-loop integral $L^{(N)}$ in Eq.~(\ref{Ln}) 
can be cast into
\beq
\label{lng}
L^{(N)}(p_1, p_2, \dots, p_N) = \int_q \;\; \prod_{i=1}^{N} \,G(q_i) \;\;,
\eeq 
where $G(q)$ denotes the customary Feynman propagator,
\beq
G(q) \equiv \frac{1}{q^2+i0} \;\;.
\eeq
We also introduce the advanced propagator $G_A(q)$,
\beq
G_A(q) \equiv \frac{1}{q^2-i0\,q_0} \;\;.
\eeq
We recall that the Feynman and advanced propagators only differ in the 
position of the particle poles in the complex plane (Fig.~\ref{fvsa}). Using 
$q^2 = q_0^2 - {\bf q}^2= 2q_+q_- - \qt^2$, we therefore have
\beq
\label{fpole}
\left[ G(q)\right]^{-1} = 0  \quad \Longrightarrow \quad
q_0 = \pm  {\sqrt {{\bf q}^2 -i0}} \;\;, {\rm or} \;\;
q_{\pm} = \frac{\qt^2-i0}{2q_{\mp}} \;\;,
\eeq
and
\beq
\left[ G_A(q)\right]^{-1} = 0 \quad \Longrightarrow \quad
q_0 \simeq \pm  {\sqrt {{\bf q}^2}} +i0 \;\;, {\rm or} \;\;
q_{\pm} \simeq \frac{\qt^2}{2q_{\mp}} +i0 \;\;.
\eeq
Thus, in the complex plane of the variable $q_0$ (or, equivalently\footnote{
	To be precise, each propagator leads to two poles in the plane $q_0$ 
	and to only one pole in the plane $q_+$ (or $q_-$).}, 
$q_{\pm}$), the pole with positive (negative) energy of the Feynman propagator is 
slightly displaced below (above) the real axis, while both poles (independently 
of the sign of the energy) of the advanced propagator are slightly 
displaced above the real axis.

\section{The Feynman theorem}
\label{sec:ft}

In this Section we briefly recall the FTT \cite{Feynman:1963ax,F2}. 

To this end, we first introduce the advanced one-loop integral $L_A^{(N)}$, which 
is obtained from $L^{(N)}$ in Eq.~(\ref{lng}) by replacing the Feynman 
propagators $G(q_i)$ with the corresponding advanced propagators $G_A(q_i)$:
\beq
\label{lna}
L_A^{(N)}(p_1, p_2, \dots, p_N) = \int_q \;\; \prod_{i=1}^{N} \,G_A(q_i) \;\;.
\eeq 
Then, we note that 
\beq
\label{lna1}
L_A^{(N)}(p_1, p_2, \dots, p_N) = 0 \;\;.
\eeq 

\begin{figure}[htb]
\begin{center}
\begin{picture}(350,130)(0,-20)
\color{blue}
\SetWidth{1.2}
\Line(0,50)(75,50)
\ArrowLine(75,50)(150,50)
\Line(75,90)(75,-15)
\ArrowArcn(75,50)(55,0,-90)
\CArc(75,50)(55,180,270)
\Text(10,85)[]{$L_A^{(N)}$}
\Text(160,40)[]{$q_0$}
\Text(135,10)[]{$C_L$}
\color{red}
\Text(110,65)[]{$\times$}
\Text(40,65)[]{$\times$}
\Text(20,65)[]{$\times$}
\Text(60,65)[]{$\times$}
\Text(90,65)[]{$\times$}
\Text(130,65)[]{$\times$}

\SetOffset(200,0)
\Line(0,50)(75,50)
\ArrowLine(75,50)(150,50)
\Line(75,90)(75,-15)
\ArrowArcn(75,50)(55,0,-90)
\CArc(75,50)(55,180,270)
\color{blue}
\Text(10,85)[]{$L^{(N)}$}
\Text(160,40)[]{$q_0$}
\Text(135,10)[]{$C_L$}
\color{red}

\Text(50,65)[]{$\times$}
\Text(110,35)[]{$\times$}
\Text(40,35)[]{$\times$}
\Text(20,65)[]{$\times$}
\Text(60,35)[]{$\times$}
\Text(90,65)[]{$\times$}

\end{picture}
\end{center}
\caption{\label{contour}
{\em Location of poles and integration contour $C_L$ in 
	the complex $q_0$-plane for the advanced (left) and Feynman (right)
	one-loop integrals, $L_A^{(N)}$ and $L^{(N)}$.  
}}
\end{figure}
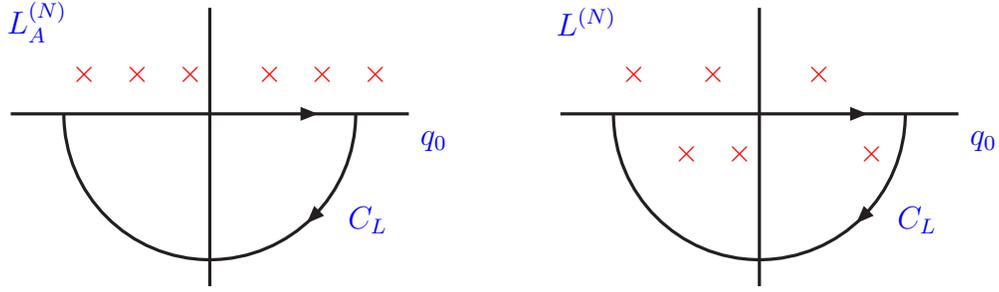

The proof of Eq.~(\ref{lna1}) can be carried out in an elementary way by 
using the Cauchy residue theorem and choosing a suitable integration path
$C_L$. We have
\beeq
\label{lna2}
\lefteqn{L_A^{(N)}(p_1, p_2, \dots, p_N) = \int_{\bf q} \;\;\; \int dq_0 \;
 \;\; \prod_{i=1}^{N} \,G_A(q_i) }\nn \\
&=& \int_{\bf q} \;\int_{C_L} dq_0 \;
 \;\; \prod_{i=1}^{N} \,G_A(q_i) 
= - \,2 \pi i \; \int_{\bf q} \;\;\sum \; \res
 \;\left[ \;\prod_{i=1}^{N} \,G_A(q_i) \right] = 0 \;\;. 
\eeeq 
The loop integral is evaluated by integrating first over the energy component 
$q_0$. Since the integrand is convergent when $q_0 \to \infty$, the $q_0$ 
integration can be performed along the contour $C_L$, which is
closed at $\infty$ in the lower half-plane of the 
complex variable $q_0$ (Fig.~\ref{contour}--{\em left}). The only 
singularities of the integrand with respect to the variable $q_0$ are 
the poles of the advanced propagators $G_A(q_i)$, which are located in the
upper half-plane. The integral along $C_L$ is then equal to 
the sum of the residues at the poles in the lower half-plane and therefore 
it vanishes. 

The advanced and Feynman propagators are related by
\beq
\label{gavsg}
G_A(q)=G(q)+\td(q) \;\;,
\eeq
which can straightforwardly be obtained by using the elementary identity 
\beq
\label{pvpid}
\frac{1}{x \pm i0} = {\rm PV}\left( \frac{1}{x} \right) \mp i \pi \, \delta(x) 
\;\;,
\eeq
where ${\rm PV}$ denotes the principal-value prescription. Inserting 
Eq.~(\ref{gavsg}) into the right-hand side of Eq.~(\ref{lna}) and collecting the 
contributions with an equal number of factors $G(q_i)$ and $\td(q_j)$, we obtain 
a relation between $L_A^{(N)}$ and the one-loop integral $L^{(N)}$:
\beeq
\label{lnavsln}
&& L_A^{(N)}(p_1, p_2, \dots, p_N) = \int_q \;\; \prod_{i=1}^{N} \,\left[
G(q_i)+\td(q_i) \right] \nn \\
&& = L^{(N)}(p_1, p_2, \dots, p_N) + L_{\rm{1-cut}}^{(N)}(p_1, p_2, \dots, p_N)
+ \dots + L_{\rm{N-cut}}^{(N)}(p_1, p_2, \dots, p_N) \;\;.
\eeeq
Here, the single-cut contribution is given by
\beq
\label{1cut}
L_{\rm{1-cut}}^{(N)}(p_1, p_2, \dots, p_N) = \int_q \;\; \sum_{i=1}^N
\; \td(q_i) \;
\prod_{\stackrel {j=1} {j\neq i}}^{N} 
\, G(q_j) \;\;.
\eeq
In general, the $m$-cut terms $L_{\rm{m-cut}}^{(N)}$ $(m \leq N)$ are the 
contributions with precisely $m$ delta functions $\td(q_i)$:
\beq
\label{lmcut}
L_{\rm{m-cut}}^{(N)}(p_1, p_2, \dots, p_N) = \int_q \;\; \left\{
\td(q_1) \dots \td(q_m) 
\; G(q_{m+1}) \dots G(q_{N}) 
+ {\rm uneq. \; perms.} \right\} \;\;,
\eeq
where the sum in the curly bracket includes all the permutations of
$q_1,\dots,q_N$ that give unequal terms in the integrand. 

Recalling that $L_A^{(N)}$ vanishes, cf.~Eq.~(\ref{lna1}), Eq.~(\ref{lnavsln})
results in:
\beq
\label{lftt}
L^{(N)}(p_1, p_2, \dots, p_N) = - \left[ \;
L_{\rm{1-cut}}^{(N)}(p_1, p_2, \dots, p_N)
+ \dots + L_{\rm{N-cut}}^{(N)}(p_1, p_2, \dots, p_N) \; \right] \;\;.
\eeq
This equation is the FTT in the specific case of the one-loop integral $L^{(N)}$. 
The FTT relates the one-loop integral $L^{(N)}$ to the multiple-cut\footnote{
	If the number of space-time dimensions is $d$, the right-hand side 
	of Eq.~(\ref{lftt}) receives contributions only from the terms with 
	$m \leq d$; the terms with larger values of $m$ vanish, since the 
	corresponding number of delta functions in the integrand is larger than 
	the number of integration variables.}
integrals $L_{\rm{m-cut}}^{(N)}$. Each delta function $\td(q_i)$ in 
$L_{\rm{m-cut}}^{(N)}$ replaces the corresponding Feynman propagator in $L^{(N)}$ 
by cutting the internal line with momentum $q_i$. This is synonymous to setting
the respective particle on shell. An $m$-particle cut decomposes the one-loop 
diagram in $m$ tree diagrams: in this sense, the FTT allows us to calculate 
loop-level diagrams from tree-level diagrams.

\begin{figure}[htb]
\vspace*{6mm}
\begin{center}
\begin{picture}(350,110)(0,-10)
\color{blue}
\SetWidth{1.2}
\BCirc(50,50){30}
\ArrowArc(50,50)(20,60,190)
\ArrowLine(39.74,78.19)(29.48,106.38)
\ArrowLine(75.98,65)(101.96,80)
\ArrowLine(69.28,27.01)(88.56,4.03)
\ArrowLine(20.45,44.79)(-9.09,39.58)
\Vertex(21.07,15.53){1.4}
\Vertex(34.60,7.71){1.4}
\Vertex(50,5){1.4}
\Text(40,110)[]{$p_1$}
\Text(115,80)[]{$p_2$}
\Text(0,30)[]{$p_N$}
\Text(100,0)[]{$p_3$}
\Text(44,55)[]{$q$}
\color{black}
\Text(-25,50)[]{$\Bigl[$}
\Text(160,50)[]{$\Bigr]_{\rm 1-cut}~~~~~\displaystyle =~~~~-~\sum_{i=1}^N$}
\color{blue}
\SetOffset(210,0)
\BCirc(50,50){30}
\ArrowArc(50,50)(30,90,130)
\ArrowArc(50,50)(30,-30,50)
\ArrowArc(50,50)(30,130,220)
\ArrowLine(69.28,72.98)(88.57,95.96)
\ArrowLine(30.71,72.98)(11.43,95.96)
\ArrowLine(75.98,35)(101.96,20)
\ArrowLine(24.02,35)(-1.96,20)
\DashLine(50,60)(50,100){5}
\Text(10,110)[]{$p_{i-1}$}
\Text(95,110)[]{$p_{i}$}
\Text(100,10)[]{$p_{i+1}$}
\Text(40,90)[]{$q$}
\Text(50,115)[]{$\tilde{\delta}(q)$}
\Text(125,55)[]{$\frac{\displaystyle 1}
{\displaystyle (q+p_i)^2+i0}$}
\Vertex(50,5){1.4}
\Vertex(65.39,7.71){1.4}
\Vertex(34.6,7.71){1.4}
\end{picture}
\end{center}
\vspace*{-4mm}
\caption{\label{fig1cut}
{\em The single-cut contribution of the Feynman Tree Theorem to the 
	one-loop $N$-point scalar integral. Graphical representation 
	as a sum of $N$ basic single-cut phase-space integrals.
}}
\end{figure}
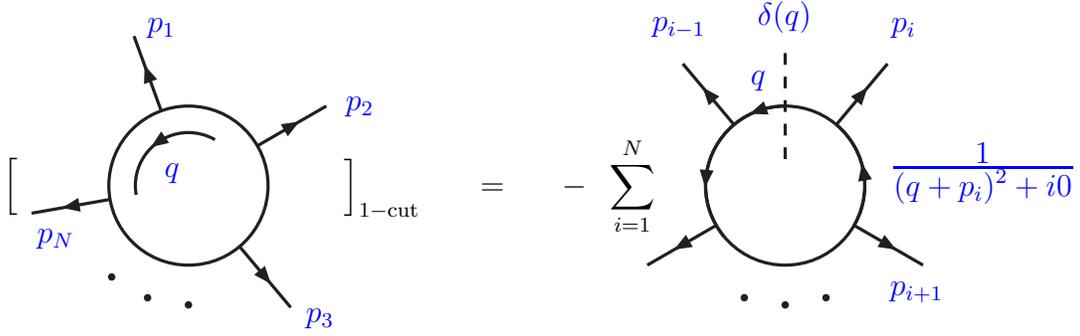

In view of the discussion in the following sections, it is useful 
to consider the single-cut contribution $L_{\rm{1-cut}}^{(N)}$ on the right-hand 
side of Eq.~(\ref{lftt}). In the case of single-cut contributions, the FTT 
replaces the one-loop integral with the customary one-particle phase-space 
integral, see Eqs.~(\ref{psm}) and (\ref{1cut}). Using the invariance of 
the loop-integration measure under translations of the loop momentum $q$,
we can perform the momentum shift $q \to q - \sum_{k=1}^i p_k$ in the term
proportional to $\td(q_i)$ on the right-hand side of Eq.~(\ref{1cut}). Thus,
cf.~Eq.~(\ref{defqi}), we have $q_i \to q$ and 
$q_j \to q + (p_{i+1} + p_{i+2} + \dots + p_{i+j})$, with $i \neq j$. We can 
repeat the same shift for each of the terms $(i=1,2,\dots,N)$ in the sum on the 
right-hand side of Eq.~(\ref{1cut}), and we can rewrite $L_{\rm{1-cut}}^{(N)}$ as 
a sum of $N$ basic phase-space integrals (Fig.~\ref{fig1cut}):
\beeq
\label{1cutsum}
L_{\rm{1-cut}}^{(N)}(p_1, p_2, \dots, p_N) &=& 
I_{\rm{1-cut}}^{(N-1)}(p_1, p_1+p_2, \dots, p_1+p_2+\dots+p_{N-1})
+ \,{\rm cyclic \;perms.} \nn \\
&=& \sum_{i=1}^N \;
I_{\rm{1-cut}}^{(N-1)}(p_i, p_i+p_{i+1}, \dots, p_i+p_{i+1}+\dots+p_{i+N-2})
\;\;.
\eeeq 
We denote the basic one-particle phase-space integrals with $n$ Feynman propagators
by $I_{\rm{1-cut}}^{(n)}$. They are defined as follows:
\beq
\label{i1cut}
I_{\rm{1-cut}}^{(n)}(k_1, k_2, \dots, k_n) = \int_q \td(q) \;
\prod_{j=1}^{n} G(q+k_j) = \int_q \td(q) \;\prod_{j=1}^{n} 
\frac{1}{2qk_j + k_j^2 + i0} \;\;.
\eeq

The extension of the FTT from the one-loop integrals $L^{(N)}$ to one-loop
scattering amplitudes ${\cal A}^{({\rm 1-loop})}$ (or Green's functions) in
perturbative field theories is straightforward, provided the corresponding
field theory is {\em unitary} and {\em local}. The generalization of 
Eq.~(\ref{lftt}) to arbitrary scattering amplitudes is \cite{Feynman:1963ax,F2}:
\beq
\label{aftt}
{\cal A}^{({\rm 1-loop})} = - \left[ \;
{\cal A}^{({\rm 1-loop})}_{\rm{1-cut}} + {\cal A}^{({\rm 1-loop})}_{\rm{2-cut}}
+ \dots \; \right] \;\;,
\eeq
where ${\cal A}^{({\rm 1-loop})}_{\rm{m-cut}}$ is obtained in the same way as
$L_{\rm{m-cut}}^{(N)}$, i.e. by starting from ${\cal A}^{({\rm 1-loop})}$
and considering all possible replacements of $m$ Feynman propagators $G(q_i)$ of 
its loop internal lines with the `cut propagators' $\td(q_i)$.

The proof of Eq.~(\ref{aftt}) directly follows from Eq.~(\ref{lftt}):
${\cal A}^{({\rm 1-loop})}$ is a linear combination of one-loop integrals that
differ from $L^{(N)}$ only by the inclusion of interaction vertices and,
eventually, particle masses. As briefly recalled below, these differences have 
harmless consequences on the derivation of the FTT.

Including particle masses in the advanced and Feynman propagators has an effect 
on the location of the poles produced by the internal lines in the loop. However, 
as long as the masses are {\em real}, as in the case of unitary theories, the 
position of the poles in the complex plane of the variable $q_0$ is affected only 
by a translation parallel to the real axis, with no effect on the imaginary part 
of the poles. This translation does not interfere with the proof of the FTT 
as given in Eqs.~(\ref{lna})--(\ref{lftt}). Therefore, the effect of a particle
mass $M_i$ in a loop internal line with momentum $q_i$ simply amounts to modifying 
the corresponding on-shell delta function $\td(q_i)$ when this line is cut to 
obtain ${\cal A}^{({\rm 1-loop})}_{\rm{m-cut}}$. This modification then leads to 
the obvious replacement:
\beq
\label{dmass}
\td(q_i) \to \td(q_i;M_i) = 2 \pi \, i \,\theta(q_{i 0}) \;\delta(q_i^2-M_i^2) 
= 2 \pi \, i \;\delta_+(q_i^2-M_i^2) \;\;.
\eeq

Including interaction vertices has the effect of introducing numerator factors
in the integrand of the one-loop integrals. As long as the theory is local,
these numerator factors are at worst polynomials of the integration momentum $q$
\footnote{This statement is not completely true in the case of gauge theories and, 
	in particular, in the case of gauge-dependent quantities. The discussion 
	of the additional issues that arise in gauge theories is postponed to 
	Sect.~\ref{sec:gauge}.} . 
In the  complex plane of the variable $q_0$, this polynomial behavior does not 
lead to additional singularities at any finite values of $q_0$. The only danger, 
when using the Cauchy theorem as in Eq.~(\ref{lna2}) to prove the FTT, stems from 
polynomials of high degree that can spoil the convergence of the $q_0$-integration 
at infinity. Nonetheless, if the field theory is unitary, these singularities at 
infinity never occur since the degree of the polynomials in the various integrands 
is always sufficiently limited by the unitarity constraint.

\section{A duality theorem}
\label{sec:dt}

In this Section we derive and illustrate the duality relation between one-loop
integrals and single-cut phase-space integrals. This relation is the main
general result of the present work.

Rather than starting from $L_A^{(N)}$, we directly apply the residue theorem 
to the computation of  $L^{(N)}$. We proceed exactly as in
Eq.~(\ref{lna2}), and obtain
\beeq
\label{ln4}
\lefteqn{L^{(N)}(p_1, p_2, \dots, p_N) = \int_{\bf q} \;\;\; \int dq_0 \;
 \;\; \prod_{i=1}^{N} \,G(q_i) }\nn\\
&=& \int_{\bf q} \;\int_{C_L} dq_0 \;
 \;\; \prod_{i=1}^{N} \,G(q_i)
= - \,2 \pi i \; \int_{\bf q} 
 \;\;\sum \; \res
 \;\left[ \;\prod_{i=1}^{N} \,G(q_i) \right] \;\;. 
\eeeq 
At variance with $G_A(q_i)$, each of the Feynman propagators $G(q_i)$ has single 
poles in both the upper and lower half-planes of the complex variable $q_0$ (see 
Fig.~\ref{contour}--{\em right}) and therefore the integral does not vanish as in 
the case of the advanced propagators. In contrast, here, the $N$ poles in the 
lower half-plane contribute to the residues in Eq.~(\ref{ln4}).

The calculation of these residues is elementary, but it involves several
subtleties. The detailed calculation, including a discussion of its subtle points, 
is presented in Appendix~\ref{app:a}. In the present Section we limit ourselves to 
sketching the derivation of the result of this computation. 
 
The sum over residues in Eq.~(\ref{ln4}) receives contributions from $N$ terms,
namely the $N$ residues at the poles with negative imaginary part of each of the 
propagators $G(q_i)$, with $i=1,\dots,N$, see Eq.~(\ref{fpole}). 
Considering the residue at the $i$-th pole we write
\beq
\label{ipole}
{\rm Res}_{\{i{\rm-th \;pole}\}}
\;\left[ \;\prod_{j=1}^{N} \,G(q_j) \right] =
\left[ {\rm Res}_{\{i{\rm-th \;pole}\}} \;G(q_i) \right]
\;\left[ \;\prod_{\stackrel {j=1} {j\neq i}}^{N}
\,G(q_j) 
\right]_{\{i{\rm-th \;pole}\}} \;\;,
\eeq
where we have used the fact that the propagators $G(q_j)$, with $j\neq i$, are not 
singular at the value of the pole of $G(q_i)$. Therefore, they can be directly 
evaluated at this value.

The calculation of the residue of $G(q_i)$ is straightforward and gives
\beq
\label{resGi}
\left[ {\rm Res}_{\{i{\rm-th \;pole}\}} \;G(q_i) \right] =
\left[ {\rm Res}_{\{i{\rm-th \;pole}\}} \;\frac{1}{q_i^2+i0} \right] 
= \int dq_0 \;\delta_+(q_i^2) \;\;.
\eeq
This result shows that considering the residue of the Feynman propagator of the 
internal line with momentum $q_i$ is equivalent to cutting that line by including the 
corresponding on-shell propagator $\delta_+(q_i^2)$. The subscript 
$+$ of $\delta_+$ refers to the on-shell mode with positive definite energy, 
$q_{i0}=|{\bf q}_i|$: the positive-energy mode is selected by the Feynman $i0$ 
prescription of the propagator $G(q_i)$. The insertion of Eq.~(\ref{resGi}) in 
Eq.~(\ref{ln4}) directly leads to a representation of the one-loop integral as a 
linear combination of $N$ single-cut phase-space integrals.

The calculation of the residue prefactor on the r.h.s. of Eq.~({\ref{ipole}) is 
more subtle (see Appendix~\ref{app:a}) and yields
\beq
\label{respre}
\left[ \;\prod_{j\neq i} \,G(q_j) \right]_{\{i{\rm-th \;pole}\}} = 
\left[ \;\prod_{j\neq i} \,\frac{1}{q_j^2 + i0} \right]_{\{i{\rm-th \;pole}\}}
 = \prod_{j\neq i} \; \frac{1}{q_j^2 - i0 \,\eta (q_j-q_i)}
\;\;,
\eeq
where $\eta$ is a {\em future-like} vector, 
\beq
\label{etadef}
\eta_\mu = (\eta_0, {\bf \eta}) \;\;, \;\; \quad \eta_0 \geq 0, 
\; \eta^2 = \eta_\mu \eta^\mu \geq 0 \;\;,
\eeq
i.e.~a $d$-dimensional vector that can be either light-like $(\eta^2=0)$ or 
time-like $(\eta^2 > 0)$ with positive definite energy $\eta_0$. Note that the 
calculation of the residue at the pole of the internal line with momentum $q_i$ 
changes the propagators of the other lines in the loop integral. Although the 
propagator of the $j$-th internal line still has the customary form $1/q_j^2$, its 
singularity at $q_j^2=0$ is regularized by a different $i0$ prescription: the 
original Feynman prescription $q_j^2 +i0$ is modified in the new prescription 
$q_j^2 - i0 \,\eta (q_j-q_i)$, which we name the `dual' $i0$ prescription or, 
briefly, the $\eta$ prescription. The dual $i0$ prescription arises from the 
fact that the original Feynman propagator $1/(q_j^2 +i0)$ is evaluated at the 
{\em complex} value of the loop momentum $q$, which is determined by the location of 
the pole at $q_i^2+i0 = 0$. The $i0$ dependence from the pole has to be combined with the 
$i0$ dependence in the Feynman propagator to obtain the total dependence as given by 
the dual $i0$ prescription. The presence of the vector $\eta_\mu$ is a consequence of using 
the residue theorem. To apply it to the calculation of the $d$ dimensional loop 
integral, we have to specify a system of coordinates (e.g.~space-time or light-cone 
coordinates) and select one of them to be integrated over at fixed values of the 
remaining $d-1$ coordinates. Introducing the auxiliary vector $\eta_\mu$ with 
space-time coordinates $\eta_\mu =(\eta_0, {\bf 0_{\perp}}, \eta_{d-1})$, 
the selected system of coordinates can be denoted in a Lorentz-invariant form.
Applying the residue theorem in the complex plane of the variable $q_0$ at fixed 
(and {\em real}) values of the coordinates $\qt$ and $q_{d-1}^\prime =
q_{d-1} - q_0 \eta_{d-1}/\eta_0$ (to be precise,
in Eq.~(\ref{ln4}) we actually used $\eta_\mu = (1, {\bf 0})$), we obtain the
result in Eq.~(\ref{respre}). 

The $\eta$ dependence of the ensuing $i0$ prescription is thus a consequence of the 
fact that the residues at each of the poles are not Lorentz-invariant quantities. 
The Lorentz-invariance of the loop integral is recovered only after summing over 
all the residues.

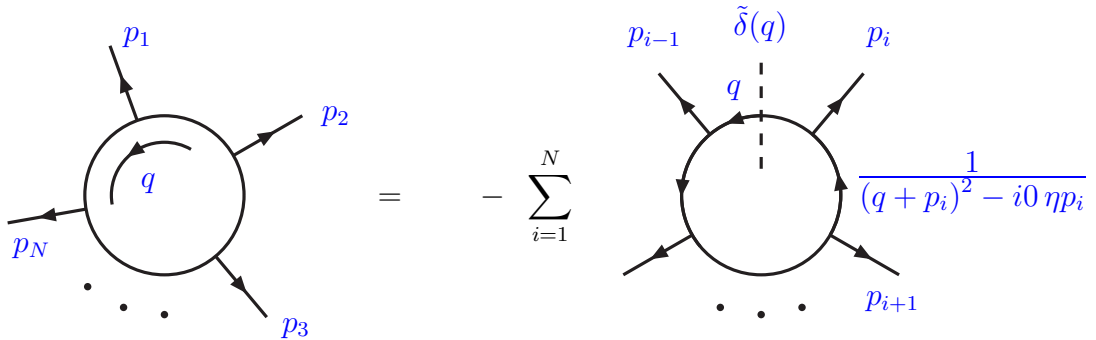
\begin{figure}[htb]
\vspace*{8mm}
\begin{center}
\begin{picture}(350,110)(0,-10)
\color{blue}
\SetWidth{1.2}
\BCirc(50,50){30}
\ArrowArc(50,50)(20,60,190)
\ArrowLine(39.74,78.19)(29.48,106.38)
\ArrowLine(75.98,65)(101.96,80)
\ArrowLine(69.28,27.01)(88.56,4.03)
\ArrowLine(20.45,44.79)(-9.09,39.58)
\Vertex(21.07,15.53){1.4}
\Vertex(34.60,7.71){1.4}
\Vertex(50,5){1.4}
\Text(40,110)[]{$p_1$}
\Text(115,80)[]{$p_2$}
\Text(0,30)[]{$p_N$}
\Text(100,0)[]{$p_3$}
\Text(44,55)[]{$q$}
\color{black}
\Text(160,50)[]{$\displaystyle =~~~~~~-~\sum_{i=1}^N$}
\color{blue}
\SetOffset(210,0)
\BCirc(50,50){30}
\ArrowArc(50,50)(30,90,130)
\ArrowArc(50,50)(30,-30,50)
\ArrowArc(50,50)(30,130,220)
\ArrowLine(69.28,72.98)(88.57,95.96)
\ArrowLine(30.71,72.98)(11.43,95.96)
\ArrowLine(75.98,35)(101.96,20)
\ArrowLine(24.02,35)(-1.96,20)
\DashLine(50,60)(50,100){5}
\Text(10,110)[]{$p_{i-1}$}
\Text(95,110)[]{$p_{i}$}
\Text(100,10)[]{$p_{i+1}$}
\Text(40,90)[]{$q$}
\Text(50,115)[]{$\tilde{\delta}(q)$}
\Text(130,55)[]{$\frac{\displaystyle 1}
{\displaystyle (q+p_i)^2-i0 \,\eta p_i}$}
\Vertex(50,5){1.4}
\Vertex(65.39,7.71){1.4}
\Vertex(34.6,7.71){1.4}
\end{picture}
\end{center}
\vspace*{-4mm}
\caption{\label{fig:dt}
{\em The duality relation for the one-loop $N$-point scalar integral.
Graphical representation as a sum of $N$ basic dual integrals.
}}
\end{figure}

Inserting the results of Eq.~(\ref{ipole})--(\ref{respre}) in Eq.~(\ref{ln4})
we directly obtain the duality relation
between one-loop integrals and phase-space integrals:
\beq
\label{ldt}
L^{(N)}(p_1, p_2, \dots, p_N) = - \; {\widetilde L}^{(N)}(p_1, p_2, \dots, p_N)
\;\;,
\eeq
where the explicit expression of the phase-space integral ${\widetilde L}^{(N)}$ 
is (Fig.~\ref{fig:dt}) 
\beq
\label{dcut}
{\widetilde L}^{(N)}(p_1, p_2, \dots, p_N) = \int_q \;\; \sum_{i=1}^N
\; \td(q_i) \;
\prod_{\stackrel {j=1} {j\neq i}}^{N} 
\; \frac{1}{q_j^2 - i0 \,\eta (q_j-q_i)}   \;\;,
\eeq
and $\eta$ is the auxiliary vector defined in Eq.~(\ref{etadef}). Each of the $N-1$ 
propagators in the integrand is regularized by the dual $i0$ prescription and, 
thus, it is named `dual' propagator. Note that the momentum difference $q_i-q_j$ 
is independent of the integration momentum $q$: it only depends on the momenta of 
the external legs of the loop (see Eq.~(\ref{defqi})).

Using the invariance of the integration measure under translations of the momentum 
$q$, we can perform the same momentum shifts as described in Sect.~\ref{sec:ft}.
In analogy to Eq.~(\ref{1cutsum}), we can rewrite Eq.~(\ref{dcut}) as a sum of $N$ 
basic phase-space integrals (Fig.~\ref{fig:dt}):
\beeq
\label{d1cutsum}
{\widetilde L}^{(N)}(p_1, p_2, \dots, p_N)
 &=& 
I^{(N-1)}(p_1, p_1+p_2, \dots, p_1+p_2+\dots+p_{N-1})
+ \,{\rm cyclic \;perms.} \nn \\
&=& \sum_{i=1}^N \;
I^{(N-1)}(p_i, p_i+p_{i+1}, \dots, p_i+p_{i+1}+\dots+p_{i+N-2})
\;\;.
\eeeq 
The basic one-particle phase-space integrals with $n$ dual propagators are denoted 
by $I^{(n)}$, and are defined as follows:
\beq
\label{idual}
I^{(n)}(k_1, k_2, \dots, k_n) =  \int_q \td(q) 
\;\; {\cal I}^{(n)}(q;k_1, k_2, \dots, k_n) =
\int_q \td(q) \;\prod_{j=1}^{n} \;
\frac{1}{2qk_j + k_j^2 - i0 \,\eta k_j} \,.
\eeq

We now comment on the comparison between the FTT (Eqs.~(\ref{1cut})--(\ref{i1cut})) 
and the duality relation (Eqs.~(\ref{ldt})--(\ref{idual})). The multiple-cut 
contributions $L_{\rm{m-cut}}^{(N)}$, with $m \geq 2$, of the FTT are completely 
absent from the duality relation, which only involves single-cut
contributions similar to those in $L_{\rm{1-cut}}^{(N)}$. However, 
the Feynman propagators present in $L_{\rm{1-cut}}^{(N)}$ are replaced by dual 
propagators in ${\widetilde L}^{(N)}$. This compensates for 
the absence of multiple-cut contributions in the duality relation. 

The $i0$ prescription of the dual propagator depends on the auxiliary vector 
$\eta$. The basic dual integrals $I^{(n)}$ are well defined for arbitrary values 
of $\eta$. However, when computing ${\widetilde L}^{(N)}$, the future-like vector 
$\eta$ has to be the {\em same} in all its contributing dual integrals 
(propagators): only then ${\widetilde L}^{(N)}$ does not depend on $\eta$.

In our derivation of the duality relation, the auxiliary vector $\eta$ originates 
from the use of the residue theorem. Independently of its origin, we can comment on 
the role of $\eta$ in the duality relation. The one-loop integral 
$L^{(N)}(p_1, p_2, \dots, p_N)$ is a function of the Lorentz-invariants $(p_i p_j)$. 
This function has a complicated analytic structure, with pole and branch-cut 
singularities (scattering singularities), in the multidimensional space of the
complex variables $(p_i p_j)$. The $i0$ prescription of the Feynman propagators
selects a Riemann sheet in this multidimensional space and, thus, it unambiguously 
defines $L^{(N)}(p_1, p_2, \dots, p_N)$ as a single-valued function. Each 
single-cut contribution to ${\widetilde L}^{(N)}$ has additional (unphysical) 
singularities in the multidimensional complex space. The dual $i0$ prescription 
fixes the position of these singularities. The auxiliary vector $\eta$ 
{\em correlates} the various single-cut contributions in ${\widetilde L}^{(N)}$, 
so that they are evaluated on the same Riemann sheet: this leads to the 
cancellation of the unphysical single-cut singularities. In contrast, in the FTT, 
this cancellation is produced by the introduction of the multiple-cut 
contributions $L_{\rm{m-cut}}^{(N)}$.

\setcounter{footnote}{0}

We remark that the expression (\ref{d1cutsum}) of ${\widetilde L}^{(N)}$ as a sum 
of basic dual integrals is just a matter of notation: for massless internal 
particles ${\widetilde L}^{(N)}$ is actually a {\em single} phase-space integral 
whose integrand is the sum of the terms obtained by cutting each of the internal 
lines of the loop. In explicit form, we can write:
\beq
\label{dlsin}
{\widetilde L}^{(N)}(p_1, p_2, \dots, p_N) =
\int_q \td(q) \;\sum_{i=1}^N \;
{\cal I}^{(N-1)}(q;p_i, p_i+p_{i+1}, \dots, p_i+p_{i+1}+\dots+p_{i+N-2})
\;\;,
\eeq
where the function ${\cal I}^{(n)}$ is the integrand of the dual integral in
Eq.~(\ref{idual}). Therefore, the duality relation (\ref{ldt}) directly expresses
the one-loop integral as the phase-space integral of a tree-level quantity. To 
name Eq.~(\ref{ldt}), we have introduced the term `duality' precisely to point out 
this direct relation\footnote{The word duality also suggests a stronger 
	(possibly one-to-one) correspondence between dual 
	integrals and loop integrals, which is further discussed in Sect.~\ref{sec:dual}.}
between the $d$-dimensional integral over the loop momentum and the
$(d-1)$-dimensional integral over the one-particle phase-space. For the 
FTT, the relation between loop-level and tree-level quantities is more involved, 
since the multiple-cut contributions $L_{\rm{m-cut}}^{(N)}$ (with $m \geq 2$) 
contain integrals of expressions that correspond to the product of $m$
tree-level diagrams over the phase-space for different number of particles.

The simpler correspondence between loops and trees in the context of the 
duality relation is further exploited in Sect.~\ref{sec:dam}, where we discuss 
Green's functions and scattering amplitudes.

\section{Example: The scalar two-point function}
\label{sec:2p}

In this Section we illustrate the application of the FTT and of the duality
relation to the evaluation of the one-loop two-point function $L^{(2)}$. A detailed 
discussion (including detailed results in analytic form and numerical results) 
of higher-point functions will be presented elsewhere \cite{inprep} (see also 
Refs.~\cite{meth, tanju}).

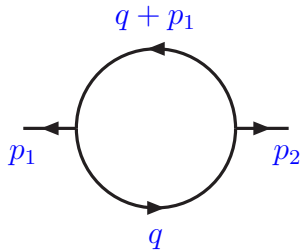
\begin{figure}[htb]
\begin{center}
\begin{picture}(100,110)(0,0)
\color{blue}
\SetWidth{1.2}
\ArrowArc(50,50)(30,0,180)
\ArrowArc(50,50)(30,180,0)
\ArrowLine(20,50)(0,50)
\ArrowLine(80,50)(100,50)
\Text(0,40)[]{$p_1$}
\Text(100,40)[]{$p_2$}
\Text(50,92)[]{$q+p_1$}
\Text(50,8)[]{$q$}

\end{picture}
\end{center}
\vspace*{-6mm}
\caption{\label{fig:bu}
{\em The one-loop two-point scalar integral $L^{(2)}(p_1,p_2)$.
}}
\end{figure}

The two-point function (Fig.~\ref{fig:bu}), also known as bubble function $\b$, 
is the simplest non-trivial one-loop integral with massless internal lines:
\beq
\label{l2}
\b (p_1^2) \equiv L^{(2)}(p_1,p_2) = - i \, 
\int \frac{d^d q}{(2\pi)^d} \;
\frac{1}{\left[ q^2+i 0 \right] \left[(q+p_1)^2+i 0 \right]} \;\;.
\eeq
Here, we have visibly implemented momentum conservation $(p_1+p_2=0)$ and exploited
Lorentz invariance ($L^{(2)}(p_1,p_2)$ can only depend on $p_1^2$, which is the
sole available invariant). Since most of the one-loop calculations have been 
carried out in four-dimensional field theories (or in their 
dimensionally-regularized versions), we set $d=4-2\ep$. Note, however, that we 
present results for arbitrary values of $\ep$ or, equivalently, for any value 
$d$ of space-time dimensions.

The result of the one-loop integral in Eq.~(\ref{l2}) is well known:
\beq
\label{bub}
\b (p^2) = c_{\Gamma} 
\;\frac{1}{\ep (1-2\ep)} \; \left( -p^2 -i 0 \right)^{-\ep} \;\;,
\eeq
where $c_{\Gamma}$ is the customary $d$-dimensional volume factor that appears
from the calculation of one-loop integrals:
\beq
c_{\Gamma} \equiv \frac{\Gamma(1+\epsilon) \; 
\Gamma^2(1-\epsilon)}{\left(4\pi\right)^{2-\epsilon}\Gamma(1-2\epsilon)} \;\;.
\eeq

We recall that the $i0$ prescription in Eq.~(\ref{bub}) follows from the
corresponding prescription of the Feynman propagators in the integrand of
Eq.~(\ref{l2}). The $i 0$ prescription defines $\b (p^2)$ as a single-value
function of the real variable $p^2$. In particular, it gives $\b (p^2)$ an
imaginary part with an unambiguous value when $p^2 > 0$:
\beq
\label{bubri}
\b (p^2) = c_{\Gamma} 
\;\frac{1}{\ep (1-2\ep)} 
\; \left( |p^2| \right)^{-\ep}
\left[ \, \theta(-p^2) + \theta(p^2) \;e^{i\pi \ep} \,
\right] \;\;.
\eeq

\subsection{General form of single-cut integrals}

To apply the FTT and the duality relation, we have to compute the single-cut 
integrals $I_{\rm{1-cut}}^{(1)}$ and $I^{(1)}$, respectively. Since these integrals 
only differ because of their $i 0$ prescription, we introduce a more general 
regularized version, $I_{\rm reg}^{(1)}$, of the single-cut integral. We define: 
\beq
\label{i1reg}
I_{\rm reg}^{(1)}(k;c(k)) = \int_q \td(q) 
\;\frac{1}{2qk + k^2 + i0 \,c(k)} = \int \frac{d^d q}{(2\pi)^{d-1}} \; 
\delta_+(q^2) \;\frac{1}{2qk + k^2 + i0 \,c(k)} \;\;.
\eeq
Although $c(k)$ is an arbitrary function of $k$, $I_{\rm reg}^{(1)}$ only depends 
on the sign of the $i0$ prescription, i.e.~on the sign of the function
$c(k)$: setting $c(k)=+1$ we recover $I_{\rm{1-cut}}^{(1)}$, cf.~Eq.~(\ref{i1cut}), 
while setting $c(k)=- \eta k$ we recover $I^{(1)}$ (see Eq.~(\ref{idual})).

The calculation of the integral in Eq.~(\ref{i1reg}) is elementary, and the
result is
\beq
\label{i1regres}
I_{\rm reg}^{(1)}(k;c(k)) = -
\frac{c_{\Gamma}}{2 \cos (\pi \ep)} 
\;\frac{1}{\ep (1-2\ep)} 
\;\left[ \frac{k^2}{k_0} - i0 \,k^2\,c(k) \right]^{-\ep}
\;\left[ k_0 - i0 \,k^2\,c(k) \right]^{-\ep}
\;\;.
\eeq
Note that the typical volume factor, ${\widetilde c}_{\Gamma}$, of the
$d$-dimensional phase-space integral is 
\beq
{\widetilde c}_{\Gamma} = \frac{\Gamma(1-\epsilon) \; 
\Gamma(1+2\epsilon)}{\left(4\pi\right)^{2-\epsilon}} \;\;.
\eeq
The factor $\cos (\pi \ep)$ in Eq.~(\ref{i1regres}) originates from the difference 
between ${\widetilde c}_{\Gamma}$ and the volume factor $c_{\Gamma}$ of the loop 
integral:
\beq
\frac{{\widetilde c}_{\Gamma}}{c_{\Gamma}} = 
\frac{\Gamma(1+2\epsilon) \, 
\Gamma(1-2\epsilon)}{\Gamma(1+\epsilon) \;\Gamma(1-\epsilon)} =
\frac{1}{\cos (\pi \ep)} \;\;.
\eeq
We also note that the result in Eq.~(\ref{i1regres}) depends on the sign of
the energy $k_0$. This follows from the fact that the integration measure
in Eq.~(\ref{i1reg}) has support on the future light-cone, which is selected
by the positive-energy requirement of the on-shell constraint 
$\delta_+(q^2)$. 

The denominator contribution $(2qk + k^2)$ in the integrand of  Eq.~(\ref{i1reg}) 
is positive definite in the kinematical region where $k^2>0$ and $k_0>0$. In this 
region the $i0$ prescription is inconsequential, and  $I_{\rm reg}^{(1)}$ has no 
imaginary part. Outside this kinematical region, $(2qk + k^2)$ can vanish, leading
to a singularity of the integrand. The singularity is regularized by the $i0$ 
prescription, which also produces a non-vanishing imaginary part. The result in 
Eq.~(\ref{i1regres}) explicitly shows these expected features, since it can be 
rewritten as 
\beeq
I_{\rm reg}^{(1)}(k;c(k)) 
&=& - 
\frac{c_{\Gamma}}{2 \cos (\pi \ep)} 
\;\frac{\left( |k^2| \right)^{-\ep}}{\ep (1-2\ep)} 
\left\{ \theta(-k^2) \left[ \cos(\pi \ep) - i \,\sin(\pi \ep) \;\sg(c(k))
\right] \right. \nn \\
\label{i1regri}
&+& \left. \theta(k^2) \left[ \theta(k_0) + \theta(-k_0)
\left( \cos(2\pi \ep) + i \,\sin(2\pi \ep) \;\sg(c(k))
\right) \right]
\right\}
\;\;.
\eeeq
We note that the functions $\b (k^2)$ and $I_{\rm reg}^{(1)}(k;c(k))$ have 
different analyticity properties in the complex $k^2$ plane. The bubble function 
has a branch-cut singularity along the positive real axis, $k^2 > 0$. The 
phase-space integral $I_{\rm reg}^{(1)}(k;c(k))$ has a branch-cut singularity 
along the entire real axis if $k_0 < 0$, while the branch-cut singularity is 
placed along the negative real axis if $k_0 > 0$.

\subsection{Duality relation for the two-point function}

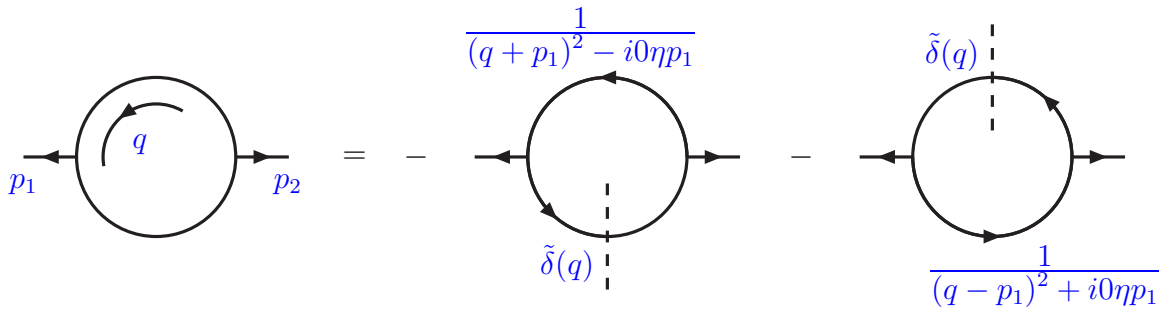
\begin{figure}[htb]
\begin{center}
\begin{picture}(410,110)(0,0)
\color{blue}
\SetWidth{1.2}
\BCirc(50,50){30}
\ArrowLine(20,50)(0,50)
\ArrowLine(80,50)(100,50)
\ArrowArc(50,50)(20,60,190)
\Text(0,40)[]{$p_1$}
\Text(100,40)[]{$p_2$}
\Text(44,55)[]{$q$}
\color{black}
\Text(125,50)[]{$=~~~$}
\color{blue}
\SetOffset(155,0)
\BCirc(50,50){30}
\ArrowArc(50,50)(30,0,180)
\ArrowArc(50,50)(30,180,270)
\ArrowLine(20,50)(0,50)
\ArrowLine(80,50)(100,50)
\DashLine(50,0)(50,40){5}
\Text(40,95)[]{$\frac{\displaystyle 1}
{\displaystyle (q + p_1)^2- i0 \eta p_1}$}
\Text(35,10)[]{$\tilde{\delta}(q)$}
\color{black}
\Text(-30,50)[]{$-$}
\color{blue}

\SetOffset(285,0)
\BCirc(50,50){30}
\ArrowArc(50,50)(30,0,90)
\ArrowArc(50,50)(30,180,0)
\ArrowLine(20,50)(0,50)
\ArrowLine(80,50)(100,50)
\DashLine(50,60)(50,100){5}
\Text(70,5)[]{$\frac{\displaystyle 1}
{\displaystyle (q - p_1)^2+i0 \eta p_1}$}
\Text(35,90)[]{$\tilde{\delta}(q)$}
\color{black}
\Text(-30,50)[]{$-$}
\color{blue}
\end{picture}
\end{center}
\caption{\label{fig:bd}
{\em One-loop two-point function: the duality relation.
}}
\end{figure}

We now consider the duality relation (Fig.~\ref{fig:bd}) in the context of this
example. The dual representation of the one-loop two-point function is given by
\beq
\label{l2dual}
{\widetilde L}^{(2)}(p_1, p_2) = 
I^{(1)}(p_1) + \Bigl( p_1 \leftrightarrow - p_1 \Bigr) \;\;,
\eeq
cf.~Eqs.~(\ref{d1cutsum}) and (\ref{idual}). The basic dual integral $I^{(1)}(k)$ 
is obtained by setting $c(k) = - \eta k$ in Eq.~(\ref{i1regres}). Since $\eta^\mu$ 
is a future-like vector, $c(k)$ has the following important property:
\beq
\label{etakey}
\sg(\eta k) = \sg(k_0) \;\;, \quad {\rm if} \;\; k^2 \geq 0 \;\;.
\eeq
Using this property, the result in Eq.~(\ref{i1regres}) can be written as 
\beq
\label{i1dual}
I^{(1)}(k) = - \;
\frac{c_{\Gamma}}{2} 
\;\frac{\left( - k^2 -i 0 \right)^{-\ep}}{\ep (1-2\ep)} 
\left[ 1 - i  \,\frac{\sin(\pi \ep)}{\cos(\pi \ep)} 
\;\sg(k^2 \eta k) \right] \;\;.
\eeq
Comparing this expression with Eq.~(\ref{bub}), we see that the imaginary 
contribution in the square bracket is responsible for the difference with the 
two-point function. However, since $\sg(- \eta k) = - \sg(\eta k)$, this 
contribution is odd under the exchange $k \to -k$ and, therefore, it cancels 
when Eq.~(\ref{i1dual}) is inserted in Eq.~(\ref{l2dual}). Taken together,
\beq
\label{l2dualr}
{\widetilde L}^{(2)}(p_1, p_2) = 
I^{(1)}(p_1) + \Bigl( p_1 \leftrightarrow - p_1 \Bigr) =
- \;
c_{\Gamma} 
\;\frac{\left( - p_1^2 -i 0 \right)^{-\ep}}{\ep (1-2\ep)} 
\;\;,
\eeq
which fully agrees with the duality relation 
${\widetilde L}^{(2)}(p_1, p_2) = - \,\b (p_1^2)$.

\subsection{FTT for the two-point function}

\begin{figure}[htb]
\begin{center}
\begin{picture}(410,210)(0,0)
\color{blue}
\SetWidth{1.2}
\SetOffset(0,100)
\BCirc(50,50){30}
\ArrowLine(20,50)(0,50)
\ArrowLine(80,50)(100,50)
\ArrowArc(50,50)(20,60,190)
\Text(0,40)[]{$p_1$}
\Text(100,40)[]{$p_2$}
\Text(44,55)[]{$q$}
\color{black}
\Text(125,50)[]{$~~=~~~$}
\color{blue}
\SetOffset(20,0)
\BCirc(50,50){30}
\ArrowArc(50,50)(30,0,180)
\ArrowArc(50,50)(30,180,270)
\ArrowLine(20,50)(0,50)
\ArrowLine(80,50)(100,50)
\DashLine(50,0)(50,40){5}
\Text(30,95)[]{$\frac{\displaystyle 1}{\displaystyle (q+p_1)^2+i0}$}
\Text(35,10)[]{$\tilde{\delta}(q)$}
\color{black}
\Text(-30,50)[]{$-$}
\color{blue}
\SetOffset(150,0)
\BCirc(50,50){30}
\ArrowArc(50,50)(30,0,90)
\ArrowArc(50,50)(30,180,0)
\ArrowLine(20,50)(0,50)
\ArrowLine(80,50)(100,50)
\DashLine(50,60)(50,100){5}
\Text(70,5)[]{$\frac{\displaystyle 1}{\displaystyle (q-p_1)^2+i0}$}
\Text(35,90)[]{$\tilde{\delta}(q)$}
\color{black}
\Text(-30,50)[]{$-$}
\color{blue}
\SetOffset(280,0)
\BCirc(50,50){30}
\ArrowArc(50,50)(30,0,90)
\ArrowArc(50,50)(30,180,270)
\ArrowLine(20,50)(0,50)
\ArrowLine(80,50)(100,50)
\DashLine(50,60)(50,100){5}
\DashLine(50,0)(50,40){5}
\Text(35,8)[]{$\tilde{\delta}(q)$}
\Text(23,90)[]{$\tilde{\delta}(q+p_1)$}
\color{black}
\Text(-30,50)[]{$-$}
\end{picture}
\end{center}
\caption{\label{fig:bftt}
{\em One-loop two-point function: the Feynman Tree Theorem
}}
\end{figure}
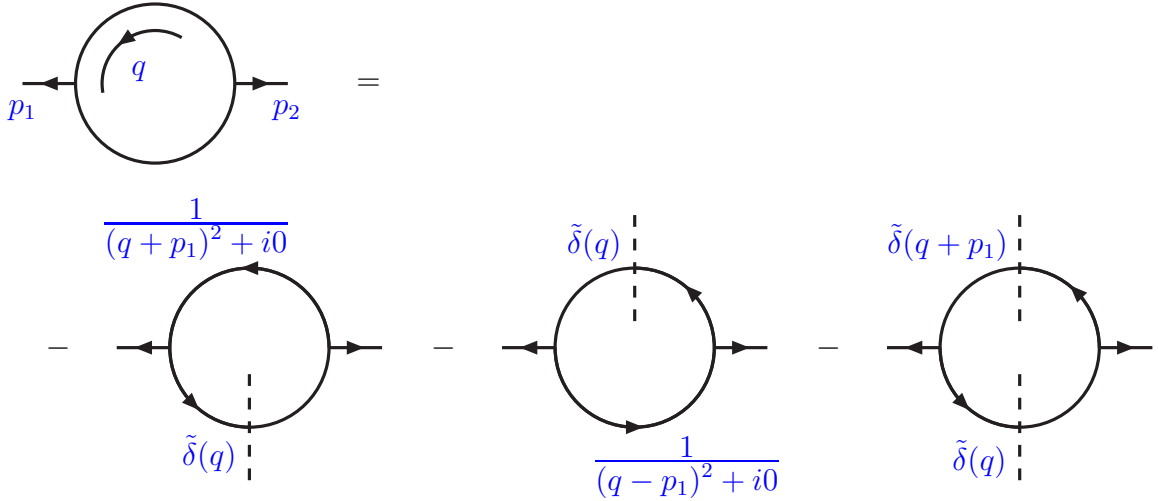

We now would like to discuss the FTT (Fig.~\ref{fig:bftt}) in the case of the 
two-point function. To this end, we want to check the relations of  
Eqs.~(\ref{lmcut})--(\ref{i1cut}). For the FTT, the two-point function is cast
into the form
\beq
\label{ftt2}
L^{(2)}(p_1, p_2, \dots, p_N) = - \left[ \;
L_{\rm{1-cut}}^{(2)}(p_1, p_2)
+ L_{\rm{2-cut}}^{(2)}(p_1, p_2) \; \right] \;\;,
\eeq
where the single-cut and double-cut contributions are 
\beq
\label{2onecut}
L_{\rm{1-cut}}^{(2)}(p_1, p_2) = 
I_{\rm{1-cut}}^{(1)}(p_1) + \Bigl( p_1 \leftrightarrow - p_1 \Bigr) \;\;,
\eeq
and
\beq
\label{2twocut}
L_{\rm{2-cut}}^{(2)}(p_1, p_2) = \int_q \;\; 
\td(q) \;\td(q+p_1) = i \, \int \frac{d^d q}{(2\pi)^{d-2}} 
\;\theta(q_0) \,\delta(q^2) \;\theta(q_0+p_{10})
\;\delta((q+p_1)^2)
 \;\;,
\eeq
respectively. The basic single-cut integral $I_{\rm{1-cut}}^{(1)}(k)$ of 
Eq.~(\ref{2onecut}) is obtained by setting $c(k) = +1$ in Eq.~(\ref{i1regres}); 
we then have
\beq
\label{i1cutr}
I_{\rm{1-cut}}^{(1)}(k) = - \;
\frac{c_{\Gamma}}{2} 
\;\frac{\left( - k^2 -i 0 \right)^{-\ep}}{\ep (1-2\ep)} 
\left[ 1 - i \, \frac{\sin(\pi \ep)}{\cos(\pi \ep)} 
\left[ \theta(-k^2) + \theta(k^2) \;\sg(k_0) \right]
\right] \;\;.
\eeq
Comparing the individual single-cut results, Eqs.~(\ref{i1dual}) and 
(\ref{i1cutr}), we see that the imaginary contributions in the square brackets 
are different. Inserting Eq.~(\ref{i1cutr}) into Eq.~(\ref{2onecut}), 
the part of the imaginary contribution that is proportional to $\sg(k_0)$ cancels 
(this part is odd under the exchange  $k \to -k$), while the remaining part does 
not:
\beq
\label{2onecutr}
L_{\rm{1-cut}}^{(2)}(p_1, p_2) = 
I_{\rm{1-cut}}^{(1)}(p_1) + \Bigl( p_1 \leftrightarrow - p_1 \Bigr) 
= - \;
c_{\Gamma}
\;\frac{\left( - p_1^2 -i 0 \right)^{-\ep}}{\ep (1-2\ep)} 
\left[ 1 - i \, \frac{\sin(\pi \ep)}{\cos(\pi \ep)} \;
\theta(-p_1^2) 
\right]  \;\;.
\eeq
We see that also the sum of the two single-cut contributions of Eqs.~(\ref{l2dualr})
and (\ref{2onecutr}) are different: the difference is due to the replacement of the
dual $i0$ prescription with the Feynman $i0$ prescription. In particular, the 
difference is a purely imaginary term with support on the space-like region 
$p_1^2 < 0$, whereas the two-point function is purely real in the same region. In 
the FTT, this difference is compensated by the double-cut contribution
$L_{\rm{2-cut}}^{(2)}$.
 
The calculation of the double-cut contribution in Eq.~(\ref{2twocut}) results in
\beq
\label{2twocutr}
L_{\rm{2-cut}}^{(2)}(p_1, p_2) = - \,i\;
c_{\Gamma}
\;\frac{\left( |p_1^2| \right)^{-\ep}}{\ep (1-2\ep)} \;
\frac{\sin(\pi \ep)}{\cos(\pi \ep)} \;\theta(-p_1^2) \;\;.
\eeq
Inserting Eqs.~(\ref{2onecutr}) and (\ref{2twocutr}) into 
the right-hand side of the FTT expression of Eq.~(\ref{ftt2}), we find
agreement with the result from
the direct one-loop computation of the two-point function.

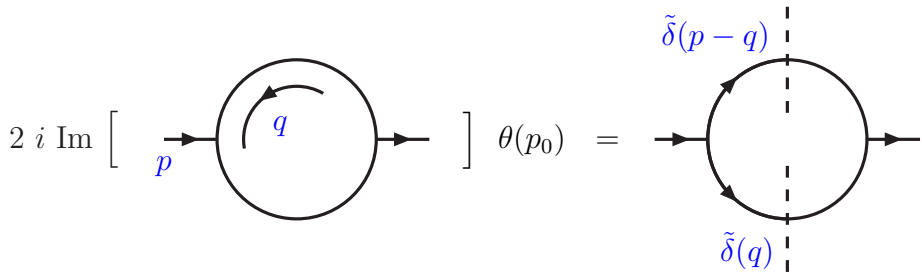
\begin{figure}[htb]
\begin{center}
\begin{picture}(300,110)(0,0)
\SetWidth{1.2}
\Text(0,50)[]{$2~i~\rm{Im}~\Bigl[ \Bigr.$}
\color{blue}
\SetOffset(30,0)
\BCirc(50,50){30}
\ArrowLine(0,50)(20,50)
\ArrowLine(80,50)(100,50)
\ArrowArc(50,50)(20,60,190)
\Text(0,40)[]{$p$}
\Text(44,55)[]{$q$}
\color{black}

\Text(125,50)[]{$~~~~~\Bigl. \Bigr]~~\theta(p_0)~~=$}
\color{blue}
\SetOffset(200,0)
\BCirc(50,50){30}
\ArrowArcn(50,50)(30,180,90)
\ArrowArc(50,50)(30,180,270)
\ArrowLine(0,50)(20,50)
\ArrowLine(80,50)(100,50)
\DashLine(50,60)(50,100){5}
\DashLine(50,0)(50,40){5}
\Text(35,8)[]{$\tilde{\delta}(q)$}
\Text(23,90)[]{$\tilde{\delta}(p-q)$}
\color{black}
\end{picture}
\end{center}
\vspace*{-3mm}
\caption{\label{fig:im}
{\em One-loop two-point function: the imaginary part.}
}
\end{figure}

To conclude this illustration of the FTT, we add a remark. The double-cut
contribution $L_{\rm{2-cut}}^{(2)}$ is different from the unitarity-cut 
contribution that gives the imaginary part of the bubble function (or, 
equivalently, the discontinuity of $\b(p^2)$ across its branch-cut singularity). 
The imaginary part of the two-point function can be obtained by applying the 
Cutkosky rules (Fig.~\ref{fig:im}):
\beq
\label{imbub}
2 \,i \;{\rm Im}\left[\b(p^2)\right] \,\theta(p_0) =
\int_q \;\; 
\td(q) \;\td(p-q) = i \, \int \frac{d^d q}{(2\pi)^{d-2}} 
\;\theta(q_0) \,\delta(q^2) \;\theta(p_{0}-q_0)
\;\delta((q-p)^2)
 \;\;.
\eeq
We see that the double-cut contributions in Eq.~(\ref{2twocut}) and (\ref{imbub}) 
are different due to the determination of the positive-energy 
flow in the internal lines. Once the energy of the line with momentum $q$ is 
fixed to be positive, the on-shell line with momentum $q+k$ has positive energy 
in Eq.~(\ref{2twocut}) and negative energy in Eq.~(\ref{imbub}). The computation 
of the double-cut integral in Eq.~(\ref{imbub}) yields
\beq
\label{imbubr}
\int_q \;\; 
\td(q) \;\td(p-q) = + \,i\;
c_{\Gamma}
\;\frac{\left( |p^2| \right)^{-\ep}}{\ep (1-2\ep)} \;2
\sin(\pi \ep) \;\theta(p^2) \;\theta(p_{0}) \;\;,
\eeq
which indeed differs from the expression in Eq.~(\ref{2twocutr}). Inserting 
Eq.~(\ref{imbubr}) in Eq.~(\ref{imbub}), we obtain the imaginary part of $\b (p^2)$,
in complete agreement with the result (\ref{bubri}) of the one-loop integral.

We also note that the Cutkosky rules in Eq.~(\ref{imbub}) can be derived in a
direct way (i.e., without the explicit computation of any integrals) from the
duality relation. The derivation is as follows. Applying the identity 
(\ref{pvpid}) to the dual propagator, we have
\beq
\label{imi1}
{\rm Im}\,[ I^{(1)}(p)] = \pi 
\int_q \;\; 
\td(q) \;\delta((q+p)^2) \;\; \sg(\eta p) \;\;.
\eeq
We now use the duality relation to compute the imaginary part of the
two-point function, which is given by
\beq
\label{imbub1}
2 \,i \;{\rm Im}\left[\b(p^2)\right] \,\theta(p_0) =
- 2\,i \;\theta(p_0) \left[ \,{\rm Im}\, I^{(1)}(p) 
+ ( p \leftrightarrow -p) \,\right] \;\;.
\eeq
Inserting Eq.~(\ref{imi1}) in Eq.~(\ref{imbub1}), we obtain
\beeq
\label{imbub2}
2 \,i \!\!\!\!\!\! \!\!\!\!\!\!\!\!
&&{\rm Im}\left[\b(p^2)\right] \,\theta(p_0) =
- 2 \pi\,i \;\sg(\eta p) \;\theta(p_0) \;
\int_q \;\;  
\td(q) \Bigl[ \,\delta((q+p)^2) - \delta((q-p)^2) \,\Bigr] \nn \\
\;&\;\; =&\! - \,(2 \pi\,i)^2 \;\sg(\eta p) \;\theta(p_0) \;
\int_q \;\;  
\delta(q^2) \;\delta((q-p)^2) \;\Bigl\{ 
\,\theta(q_0-p_0) - \theta(q_0) \,\Bigr\} \;\;,
\eeeq
where the first term in the square bracket has been rewritten by performing the
shift $q \to q-p$ of the integration variable $q$. The energy constraints in 
Eq.~(\ref{imbub2}) result in
\beq
\theta(p_0) \;\Bigl\{ 
\,\theta(q_0-p_0) - \theta(q_0) \,\Bigr\} = - \,\theta(q_0) \;\theta(p_0-q_0)
\;\;.
\eeq
This can be inserted in Eq.~(\ref{imbub2}) to obtain
\beq
\label{imbub3}
2 \,i \;{\rm Im}\left[\b(p^2)\right] \,\theta(p_0) = \sg(\eta p) \;
\int_q \;\;\td(q) \;\td(p-q) \;\;.
\eeq
We observe that the constraints $q^2=(p-q)^2=0$ and $q_0 > 0, \,p_0-q_0 > 0$ 
imply $\sg(\eta q)=\sg(\eta (p-q)) = +1$ (see Eq.~(\ref{etakey})) and, hence,
$\sg(\eta p)=+1$. Therefore Eq.~(\ref{imbub3}) becomes identical to 
Eq.~(\ref{imbub}).

\section{Relating Feynman's theorem and the duality theorem}
\label{sec:rel}

The one-loop integral $L^{(N)}$ can be expressed by using either the FTT or the 
duality relation. Comparing Eq.~(\ref{lftt}) with Eq.~(\ref{ldt}), we thus derive
\beq
\label{fttvsdt}
{\widetilde L}^{(N)}(p_1, p_2, \dots, p_N)
= 
L_{\rm{1-cut}}^{(N)}(p_1, p_2, \dots, p_N)
+ \dots + L_{\rm{N-cut}}^{(N)}(p_1, p_2, \dots, p_N) \;\;.
\eeq
This expression relates single-cut dual integrals with multiple-cut Feynman
integrals. It has been derived in an indirect way, by applying the residue
theorem to the evaluation of one-loop integrals.

In this Section we present another proof of Eq.~(\ref{fttvsdt}). The proof
is direct and purely algebraic. It further clarifies the connection between the 
FTT and the duality relation.

Our starting point is a basic identity between dual and Feynman propagators.
The identity applies to the dual propagators when they are inserted in a
single-cut integral. Then
\beeq
\td(q) \;\frac{1}{2qk + k^2 - i0 \,\eta k} &=&
\td(q) \;\Bigl[ G(q+k) + \theta(\eta k) \;2\pi i \;\delta((q+k)^2) 
\Bigr] \nn \\
\label{dovsfp}
&=& \td(q) \;\Bigl[ G(q+k) + \theta(\eta k) \;\td(q+k) \Bigr] \;\;.
\eeeq
The equality on the first line of Eq.~(\ref{dovsfp}) directly follows from
Eq.~(\ref{pvpid}). The equality on the second line is obtained as follows. 
Using the constraint $\td(q)$, we have $q^2=0$ and $q_0>0$. Therefore, from 
Eq.~(\ref{etakey}) we thus have $\eta q >0$. Using $\eta q >0$ and the constraint 
$\theta(\eta k)$, we have $\eta (q+k) >0$. Combining $\eta (q+k) >0$ with 
$(q+k)^2 =0$, from Eq.~(\ref{etakey}) we thus have $q_0+k_0>0$. This enables the 
replacement $\delta((q+k)^2) \to \delta_+((q+k)^2)$, which finally yields 
Eq.~(\ref{dovsfp}).

\subsection{Two-point function}

The relation (\ref{dovsfp}) can be used to prove Eq.~(\ref{fttvsdt}). We first 
consider the case $N=2$. Inserting Eq.~(\ref{dovsfp}) in Eq.~(\ref{idual}) and 
comparing with Eqs.~(\ref{i1cut}) and (\ref{2twocut}), we obtain
\beq
I^{(1)}(p_1) = I_{\rm{1-cut}}^{(1)}(p_1) + \theta(\eta p_1)
\int_q \td(q) \;\td(q+p_1)
= I_{\rm{1-cut}}^{(1)}(p_1) + \theta(\eta p_1) 
\;L_{\rm{2-cut}}^{(2)}(p_1, p_2) \;\;.
\eeq
We can now use this equation to compute ${\widetilde L}^{(2)}$:
\beq
{\widetilde L}^{(2)}(p_1, p_2) = I^{(1)}(p_1) + I^{(1)}(p_2)
= L_{\rm{1-cut}}^{(2)}(p_1, p_2) + 
\Bigl[ \theta(\eta p_1) + \theta(\eta p_2) \Bigr]
L_{\rm{2-cut}}^{(2)}(p_1, p_2) \;\;.
\eeq
This relation is equivalent to Eq.~(\ref{fttvsdt}), since by merely using
momentum conservation, $p_1+p_2=0$, we find
\beq
\theta(\eta p_1) + \theta(\eta p_2) = \theta(\eta p_1) + \theta(- \eta p_1)
= 1 \;\;.
\eeq

\subsection{General $N$-point function}

More generally, the identity (\ref{dovsfp}) relates the basic dual integrals
$I^{(n)}$ with multiple-cut Feynman integrals. Inserting Eq.~(\ref{dovsfp})
in Eq.~(\ref{idual}) and using Eq.~(\ref{i1cut}), we obtain
\beeq
\label{idualvseta}
I^{(n)}(k_1, k_2, \dots, k_n) &=& I_{\rm{1-cut}}^{(n)}(k_1, k_2, \dots, k_n)
+ I_{\eta}^{(n)}(k_1, k_2, \dots, k_n) \nn \\
&=& I_{\rm{1-cut}}^{(n)}(k_1, k_2, \dots, k_n) + \sum_{m=1}^{n}
I_{m, \eta}^{(n)}(k_1, k_2, \dots, k_n) \;\;,
\eeeq
where
\beeq
\label{metacut}
I_{m, \eta}^{(n)}(k_1, k_2, \dots, k_n) = \int_q \;\td(q)
\!\!\!\!\!\!\!\! &&\Bigl\{
\td(q+k_1) \dots \td(q+k_m) 
\; G(q+k_{m+1}) \dots G(q+k_{n}) \Bigr. \nn \\
&& \; \times \; \theta(\eta k_1) \dots \theta(\eta k_m) + 
\; \Bigl. {\rm uneq. \; perms.} \Bigr\} \;\;.
\eeeq
Note that the key difference between $I_{m, \eta}^{(n)}$ and the multiple-cut
contributions of the FTT (see Eq.~(\ref{lmcut})) is the presence of 
the momentum constraints, $\theta(\eta k_i)$, in Eq.~(\ref{metacut}). 


For a proof in the case of the $N$-point function, we employ the following 
relation: 
\beq
\label{metavsmcut}
I_{m-1, \eta}^{(N-1)}(p_1, p_1+p_2, \dots, p_1+p_2+\dots+p_{N-1})
+ \,{\rm cyclic \;perms.} = L_{\rm{m-cut}}^{(N)}(p_1, p_2, \dots, p_N) \;\;.
\eeq	
Summing over the cyclic permutations of $I^{(N-1)}$ as in Eq.~(\ref{d1cutsum}), and
using Eqs.~(\ref{idualvseta}), (\ref{1cutsum}) and (\ref{metavsmcut}), 
we straightforwardly obtain the relation in  Eq.~(\ref{fttvsdt}). 

We note that the proof of Eq.~(\ref{metavsmcut}) is mainly a matter of 
combinatorics, and it does not require the explicit evaluation of any $m$-cut 
integral. Eventually, the main ingredient of the proof is the following 
algebraic identity
\beq
\label{thetacon}
\theta(\eta p_1) \,\theta(\eta (p_1+p_2)) \,\dots 
\,\theta(\eta (p_1+p_2+ \dots+p_{N-1})) 
 + {\rm cyclic \;\; perms.} = 1 \;\;.
\eeq
It is a direct consequence of momentum conservation, namely  $\sum_{i=1}^N p_i=0$. 
The derivation of Eq.~(\ref{thetacon}) is presented in Appendix~\ref{app:b}.

To simplify the combinatorics in the proof of Eq.~(\ref{metavsmcut}), we first 
rewrite $I_{m, \eta}^{(n)}$ in Eq.~(\ref{metacut}) as
\beq
\label{rmetacut}
I_{m, \eta}^{(n)}(k_1, k_2, \dots, k_n) = 
I_{m, F}^{(n)}(k_1, k_2, \dots, k_n) + 
\delta I_{m, \eta}^{(n)}(k_1, k_2, \dots, k_n) \;\;,
\eeq
where
\beeq
\label{lmfcut}
I_{m, F}^{(n)}(k_1, k_2, \dots, k_n) = \frac{1}{m+1} \;\int_q \;\td(q)
\!\!\!\!\!\!\!\! &&\Bigl\{
\td(q+k_1) \dots \td(q+k_m) 
\; G(q+k_{m+1}) \dots G(q+k_{n}) \Bigr. \nn \\
&& \; + \; \Bigl. {\rm uneq. \; perms.} \Bigr\} \;\;,
\eeeq
and
\beeq
\label{lmdcut}
\delta I_{m, \eta}^{(n)}(k_1, k_2, \dots, k_n) = \int_q \;\td(q)
\!\!\!\!\!\!\!\!\! &&\left\{
\td(q+k_1) \dots \td(q+k_m) 
\; G(q+k_{m+1}) \dots G(q+k_{n}) \frac{}{} \right. \nn \\
&& \, \times \; \left. \left[
\theta(\eta k_1) \dots \theta(\eta k_m) - \frac{1}{m+1} \right] \; + \,
{\rm uneq. \; perms.} \right\} .
\eeeq
This leaves us with the task to prove the relations
\beq
\label{mfvsmcut}
I_{m-1, F}^{(N-1)}(p_1, p_1+p_2, \dots, p_1+p_2+\dots+p_{N-1})
+ \,{\rm cyclic \;perms.} = L_{\rm{m-cut}}^{(N)}(p_1, p_2, \dots, p_N) \;\;,
\eeq
and
\beq
\label{dmetavsmcut}
\delta I_{m-1, \eta}^{(N-1)}(p_1, p_1+p_2, \dots, p_1+p_2+\dots+p_{N-1})
+ \,{\rm cyclic \;perms.} = 0  \;\;.
\eeq
Obviously, Eqs.~(\ref{rmetacut}), (\ref{mfvsmcut}) and (\ref{dmetavsmcut}) imply 
Eq.~(\ref{metavsmcut}).

The relation (\ref{mfvsmcut}) can be proven as follows. According to
Eq.~(\ref{lmcut}), $L_{\rm{m-cut}}^{(N)}$ is a sum of $m$-cut contributions with a 
fully symmetric dependence on the momenta $q_i$ of the internal lines of the loop 
integral. The expression on the left-hand side of Eq.~(\ref{mfvsmcut}) is also a 
fully symmetric linear combination of $m$-cut contributions: the symmetrization 
follows from the sum over the permutations in Eqs.~(\ref{lmfcut}) and 
(\ref{mfvsmcut}). Hence, owing to their symmetry, the left-hand side and the right-hand 
side of Eq.~(\ref{mfvsmcut}) are necessarily proportional, and the 
proportionality coefficient is just unity. To show this, we can give weight unity 
to each $m$-cut contribution and simply count the number of $m$-cut 
contributions on both sides of Eq.~(\ref{mfvsmcut}). The number of terms in 
$L_{\rm{m-cut}}^{(N)}$ equals the total number of permutations in the curly bracket 
of Eq.~(\ref{lmcut}), namely
\beq
\label{numb1}
\binom{ \,N\, }{ \,m\, } = \frac{N!}{m! \;(N-m)!} \;\;.
\eeq
The number of terms on the left-hand side of Eq.~(\ref{mfvsmcut}) is
\beq
\label{numb2}
\frac{1}{m} \;
\binom{ \,N-1\, }{ \,m-1\, } \;N = \frac{1}{m} \;\frac{(N-1)!}{(m-1)! \;(N-m)!}
\;N \;\;,
\eeq
where the factor $1/m$ is the weight of each contribution 
to $I_{m-1, F}^{(N-1)}$, the factor $\binom{ \,N-1\, }{ \,m-1\, }$ is the number 
of permutations that contribute to $I_{m-1, F}^{(N-1)}$ (see Eq.~(\ref{lmfcut})), 
and the factor $N$ is the number of cyclic permutations in Eq.~(\ref{mfvsmcut}).
As we can see, the numbers given by Eqs.~(\ref{numb1}) and (\ref{numb2}) coincide,
thus yielding the equality in Eq.~(\ref{mfvsmcut}).

The relation (\ref{dmetavsmcut}) can be proven as follows. The left-hand side
is a sum of $m$-cut contributions of the loop integral $L^{(N)}$. We can
organize these contributions in a sum of $\binom{ \,N\, }{ \,m\, }$ diagrams
as on the right-hand side of Eq.~(\ref{lmcut}): each diagram has $m$ fixed
internal lines that have been cut. The coefficient of each diagram is computed
according to the expression on the left-hand side of Eq.~(\ref{dmetavsmcut}).
As discussed below, this coefficient vanishes algebraically, thus yielding the
result in Eq.~(\ref{dmetavsmcut}).

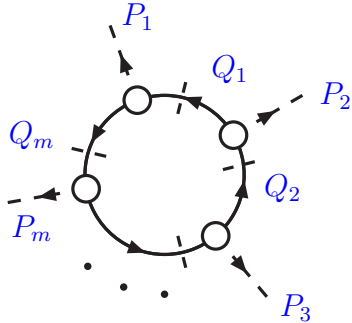
\begin{figure}[htb]
\begin{center}
\vspace*{5mm}
\begin{picture}(120,110)(0,-10)
\color{blue}
\SetWidth{1.2}
\BCirc(50,50){30}
\ArrowArc(50,50)(30,110,190)
\ArrowArc(50,50)(30,190,-50)
\ArrowArc(50,50)(30,-50,30)
\ArrowArc(50,50)(30,30,110)
\DashArrowLine(39.74,78.19)(29.48,106.38){4}
\DashArrowLine(75.98,65)(101.96,80){4}
\DashArrowLine(69.28,27.01)(88.56,4.03){4}
\DashArrowLine(20.45,44.79)(-9.09,39.58){4}
\BCirc(39.74,78.19){5}
\BCirc(75.98,65){5}
\BCirc(69.28,27.01){5}
\BCirc(20.45,44.79){5}
\DashLine(55,73)(58,85){5}
\DashLine(72,52)(84,55){5}
\DashLine(55,27)(58,15){5}
\DashLine(15,60)(28,57){5}

\Vertex(21.07,15.53){1.4}
\Vertex(34.60,7.71){1.4}
\Vertex(50,5){1.4}
\Text(40,110)[]{$P_1$}
\Text(75,90)[]{$Q_1$}
\Text(115,80)[]{$P_2$}
\Text(95,45)[]{$Q_2$}
\Text(0,65)[]{$Q_m$}
\Text(0,30)[]{$P_m$}
\Text(100,0)[]{$P_3$}
\end{picture}
\end{center}
\vspace*{-4mm}
\caption{\label{mcut}
{\em A one-loop diagram with $m$ cut lines. Each blob denotes a set of internal
lines that are not cut.
}}
\end{figure}

We consider one of the diagram with $m$ cut lines, and we denote the momenta of
these internal lines as $Q_1, Q_2, \dots, Q_m$ (Fig.~\ref{mcut}). We define
$P_i= Q_i - Q_{i-1}$, so that $P_i$ is the total external momentum 
between the cut lines with momenta 
$Q_i$ and $Q_{i-1}$. The computation of the diagram involves the factor
\beq
\label{dtm}
\td(Q_1) \;\td(Q_2) \,\dots\,  \td(Q_m) \;\;,
\eeq
and two other factors. One factor stems from the product of the Feynman propagators 
of the uncut internal lines and it is inconsequential to the 
present discussion. The other factor arises from the term in the square bracket on 
the right-hand side of Eq.~(\ref{lmdcut}). We note that 
$\delta I_{m-1, \eta}^{(N-1)}$ involves the product 
$\td(q) \;\td(q+k_1) \,\dots\,  \td(q+k_{m-1})$ of $m$ delta functions, but the 
term in the square bracket is symmetric only with respect to the argument of 
$m-1$ delta functions. Therefore, inserting Eq.~(\ref{lmdcut}) into 
Eq.~(\ref{dmetavsmcut}) and performing the sum over the permutations, the term in 
the square bracket leads to $m$ different contributions: each contribution 
corresponds to one of the assignments $\td(q) \to \td(Q_i)$ with $i=1,2,\dots,m$. 
In conclusion, the square-bracket term contributes to multiply the left-hand side 
of Eq.~(\ref{dtm}) by a factor proportional to the following expression:
\beeq
&& \!\!\!\!\!\!\!\!\!\!\!\!\!\!\!\!
\left[ 
\theta(\eta P_1) \,\theta(\eta (P_1+P_2)) \,\dots 
\,\theta(\eta (P_1+P_2+ \dots+P_{m-1})) - \frac{1}{m} \,
\right] + {\rm cyclic \;\; perms.} \nn \\
&&\!\!\!\!\!\!\!\!\!\!\!\!\!\!
= \Bigl\{ \theta(\eta P_1) \,\theta(\eta (P_1+P_2)) \,\dots 
\,\theta(\eta (P_1+P_2+ \dots+P_{m-1})) + {\rm cyclic \;\; perms.}
\Bigr\} - 1 \;\;.
\eeeq
This expression vanishes, because of Eq.~(\ref{thetacon}) and the 
momentum conservation constraint $\sum_{i=1}^m P_i=0$. Therefore, each  
$m$-cut diagram of the left-hand side of Eq.~(\ref{dmetavsmcut}) 
has a vanishing coefficient.

\section{Dual bases and generalized duality}
\label{sec:dual}

One-loop Feynman integrals and single-cut dual integrals are not in a
{\em one-to-one} correspondence. Starting from this observation we discuss in more
general terms the nature of the correspondence between one-loop and single-cut 
integrals in this section.

Using the duality relation, any one-loop Feynman integral $L^{(N)}$ can be 
expressed as a linear combination of the basic dual integrals $I^{(N-1)}$, but the 
opposite statement is not true. Therefore, the dual integrals $I^{(n)}$ form a 
linear basis of the functional space generated by the loop integrals, but overall
they generate a larger space containing that of the one-loop Feynman 
integrals.

To express $I^{(N-1)}$ as a linear combination of loop integrals, we have to
introduce generalized one-loop integrals, whose integrands contain both Feynman 
and advanced propagators. We define them through
\beq
\label{Lng}
L^{(N)}(p_1,\alpha_1, p_2, \alpha_2, \dots, p_N, \alpha_N) =  
\int_q  \;\,\prod_{i=1}^{N} \, G_{\alpha_i}(q_i) \;\;,
\eeq
where the label $\alpha_i$ can take two values, $\alpha_i=F,A$, and 
$G_{F}(q_i)=G(q_i)$ is the Feynman propagator, while $G_{A}(q_i)$ is the advanced 
propagator. In particular, when $\alpha_1=\alpha_2=\dots=\alpha_N=F$ we recover 
the one-loop Feynman integral in Eq.~(\ref{lng}), while we obtain  the one-loop 
advanced integral in Eqs.~(\ref{lna}) and (\ref{lna1}) for the case 
$\alpha_1=\alpha_2=\dots=\alpha_N=A$.

The relation between $I^{(N-1)}$ and the generalized one-loop integrals in 
Eq.~(\ref{Lng}) is obtained by rewriting the dual propagators 
as a linear combination of $G$ and $G_{A}$. Using Eqs.~(\ref{gavsg}) and 
(\ref{dovsfp}) we have:
\beeq
\td(q) \;\frac{1}{2qk + k^2 - i0 \,\eta k} &=&
\td(q) \;\Bigl[ G(q+k) + \theta(\eta k) \Bigl( G_A(q+k)- G(q+k)\Bigr) 
\; \Bigr] \nn \\
\label{dovsfap}
&=& \td(q) \;\Bigl[ \,\theta(-\eta k) \,G(q+k) 
+ \theta(\eta k) \,G_A(q+k) \,\Bigr]
\;\;,
\eeeq
which can be inserted in Eq.~(\ref{idual}). We thus obtain
\beeq
I^{(n)}(k_1, k_2, \dots, k_n) \!\!\!\!\!\!\!\!&&= 
\int_q \td(q) \;\prod_{j=1}^{n} \;
\Bigl[ \,\theta(-\eta k_j) \,G(q+k_j) 
+ \theta(\eta k_j) \,G_A(q+k_j) \,\Bigr] \nn \\
\label{dualvsloop}
=&&\!\!\!\!\!\!\!\!\! \int_q \;\Bigl( G_A(q)- G(q) \Bigr) \;\prod_{j=1}^{n} \;
\Bigl[ \,\theta(-\eta k_j) \,G(q+k_j) 
+ \theta(\eta k_j) \,G_A(q+k_j) \,\Bigr]
\,,
\eeeq
where again we have used Eq.~(\ref{gavsg}) to express $\td(q)$ as a linear
combination of $G(q)$ and $G_{A}(q)$. The right-hand side of Eq.~(\ref{dualvsloop}) 
is a sum of generalized one-loop integrals. Note that the $\eta$ dependence of 
$I^{(n)}$ appears only in the coefficients $\theta(\pm \eta k_j)$.

In the simplest case, with $n=1$, Eq.~(\ref{dualvsloop}) reads:
\beeq
I^{(1)}(p_1)&=& -\; \theta(-\eta p_1) \int_q \;G(q) \;G(q+p_1) \nn \\
&+& \left[ 
\theta(-\eta p_1) \int_q \;G_A(q) \;G(q+p_1) -
\theta(\eta p_1) \int_q \;G(q) \;G_A(q+p_1) \right]  \\
&=& -\; \theta(-\eta p_1) \;L^{(2)}(p_1,-p_1) + \Bigl[ 
\theta(-\eta p_1) \;L^{(2)}(p_1,F, -p_1,A) - \left( p_1 \leftrightarrow 
-p_1 \right)
\Bigr] \;\;, \nn
\eeeq
where we have used Eq.~(\ref{lna1}). Note that the term in the square bracket is
odd under the exchange $p_1 \leftrightarrow -p_1$. Therefore the sum 
$I^{(1)}(p_1) + I^{(1)}(-p_1)$ consistently reproduces the duality relation
(i.e., equivalently, it reproduces the two-point function $L^{(2)}(p_1,-p_1)$).

More generally, the linear relation in Eq.~(\ref{dualvsloop}) implies that the
dual integrals $I^{(N-1)}$ belong to the functional space
that is generated by
the generalized one-loop integrals of Eq.~(\ref{Lng})

Nonetheless, we have not yet established a one-to-one 
correspondence between single-cut and one-loop integrals. In fact, the 
correspondence in Eq.~(\ref{dualvsloop}) is not invertible.  The generalized 
one-loop integrals can be expressed in terms of single-cut integrals by a proper
generalization of the duality relation in Eqs.~(\ref{ldt}) and (\ref{dcut}).
However, the single-cut integrals of this generalized relation involve 
the integration of both dual and advanced propagators.

The generalized duality relation is:
\beeq
L^{(N)}(p_1,\alpha_1, p_2, \alpha_2, \dots, p_N, \alpha_N) \!\!
&=& - \int_q \;\; \sum_{i=1}^N 
\; \td(q_i) \;\delta_{\alpha_i,F} \nn \\
\label{Lngvscut}
\!\!&\times&
\prod_{\stackrel {j=1} {j\neq i}}^{N} \left[ \delta_{\alpha_j,F} \;
\; \frac{1}{q_j^2 - i0 \,\eta (q_j-q_i)}  
+ \delta_{\alpha_j,A} \;G_{A}(q_j)
\right] \,.
\eeeq
This result can be derived by applying the residue theorem 
(see Appendix~\ref{app:a}).

Alternatively, Eq.~(\ref{Lngvscut}) can also be derived by applying an algebraic 
procedure similar to the one used in Sect.~\ref{sec:rel} to prove 
Eq.~(\ref{fttvsdt}). This procedure consists of rewriting  the right-hand side of 
Eqs.~(\ref{Lng}) and (\ref{Lngvscut}) as multiple-cut integrals of expressions 
involving only {\em advanced} propagators. The resulting expressions can be shown 
to agree with each other. The rewrite of Eqs.~(\ref{Lng}) 
and (\ref{Lngvscut}) is achieved by using Eq.~(\ref{gavsg}) to replace Feynman and 
dual propagators with advanced propagators. More precisely, in the case of the 
dual propagators, Eqs.~(\ref{gavsg}) and (\ref{dovsfap}) give: 
\beq
\label{dovsap}
\td(q) \;\frac{1}{2qk + k^2 - i0 \,\eta k} =
\td(q) \;\Bigl[ G_A(q+k) - \theta(-\eta k) \;\td(q+k)
 \; \Bigr]
\;\;.
\eeq

To exemplify this algebraic procedure, we can explicitly show its application
to the simple, though non-trivial, case of the one-loop integral 
$L^{(3)}(p_1, F, p_2, F, p_3, A)$. The right-hand side of Eq.~(\ref{Lng})
yields
\beeq
&&\!\!\!\!\!\!\!\!\!\!\!\!\!\!\!\!\!\! \int_q \;G_A(q) \;G(q+p_1) 
\;G(q+p_1+p_2) =  
- \int_q \;G_A(q) \nn \\
\label{exl}
&&\!\!\!\!\!\!\!\!\! \!\!\!\!\!\!\!\!\!
\times
 \Bigl[ \td(q+p_1) \;G_A(q+p_1+p_2)
+ \td(q+p_1+p_2) \;G_A(q+p_1) - \td(q+p_1) \;\td(q+p_1+p_2) \Bigr] \,,
\eeeq
where we have also used Eq.~(\ref{lna1}). After using Eq.~(\ref{dovsap}), the 
right-hand side of Eq.~(\ref{Lngvscut}) reads
\beeq
- \int_q \;G_A(q) &&\!\!\!\!\!\!\!\!\!
\left[ \,\td(q+p_1) \,\frac{1}{(q+p_1+p_2)^2 -i 0 \,\eta p_2}
+\td(q+p_1+p_2) \,\frac{1}{(q+p_1)^2 +i 0 \,\eta p_2} \,\right] \nn \\
\label{exd}
= &&\!\!\!\!\!\! - \int_q \;G_A(q) \;\Bigl[ \td(q+p_1) \Bigl( G_A(q+p_1+p_2) -
\theta(-\eta p_2) \;\td(q+p_1+p_2) \Bigr) \Bigr. \nn \\
&&\;\;\;\;\;\;\;\;\;\;\;\;\;\;\;\;\; + \;\Bigl. \td(q+p_1+p_2) 
\Bigl( G_A(q+p_1) 
- \theta(\eta p_2) \;\td(q+p_1)
\Bigr) \Bigr] \;.
\eeeq
By simple inspection, we see that the expressions in Eqs.~(\ref{exl}) and 
(\ref{exd}) coincide.

The generalized duality in Eq.~(\ref{Lngvscut}) relates one-loop integrals to 
single-cut phase-space integrals. Note that only the Feynman propagators of the 
loop integral are cut; the uncut Feynman propagators are instead 
replaced by dual propagators. The advanced propagators of the loop integral are 
not cut, and they appear unchanged in the integrand of the phase-space integral.

Moreover, the correspondence in Eq.~(\ref{Lngvscut}) between one-loop and single-cut 
integrals is invertible. Using the same algebraic steps as in Eqs.~(\ref{dovsfap}) 
and (\ref{dualvsloop}), we indeed obtain:
\beeq
&&\!\!\!\!\!\!\!\!\!\!\!\!\!\!
\int_q \;\td(q) \left( \prod_{j=1}^{m} \frac{1}{2qk_j+k_j^2 - i0 \,\eta k_j}
\right)
\prod_{i=1}^{k} G_A(q+k_i)
\nn \\
\label{dinv}
&&\!\!\!\!\!\!\!\!\!\!\!\!\!\!\!
= \int_q \;\Bigl( G_A(q)- G(q) \Bigr) \;\prod_{j=1}^{m} \;
\Bigl[ \,\theta(-\eta k_j) \,G(q+k_j) 
+ \theta(\eta k_j) \,G_A(q+k_j) \,\Bigr]
\prod_{i=1}^{k} G_A(q+k_i) \;.
\eeeq 
The functional space generated by the generalized one-loop integrals is thus
equivalent to the space generated by the single-cut integrals on the left-hand
side of Eq.~(\ref{dinv}). The one-loop integrals of Feynman and advanced
propagators and the single-cut integrals of dual and advanced propagators
can therefore be regarded as equivalent dual basis of the same functional space.

\setcounter{footnote}{1}
\renewcommand{\thefootnote}{\fnsymbol{footnote}}

\section{Massive integrals, complex masses and unstable \\
	particles}
\label{sec:mass}

As discussed at the end of Sect.~\ref{sec:ft}, the introduction of particle masses
and massive propagators does not lead to difficulties in the generalization of 
the FTT from the massless case. The same discussion and the same conclusions apply 
to the duality relation, since this relation can be derived by applying the 
residue theorem in close analogy with the derivation of the FTT. Therefore, as 
long as the mass is {\em real}, the effect of a particle mass $M_i$ in the Feynman 
propagator of a loop internal line with momentum $q_i$ amounts to modifying 
(according to the replacement in Eq.~(\ref{dmass})) the corresponding on-shell 
delta function $\td(q_i)$ when this line is cut to obtain the dual representation 
${\widetilde L}^{(N)}$ (see Eqs.~(\ref{dcut}) and (\ref{Lngvscut})) of the loop 
integral $L^{(N)}$. Note also that the $i0$ prescription of the dual propagators 
is not affected by the masses. More precisely, if the Feynman propagator of the 
$j$-th internal line has mass $M_j$, the corresponding dual propagator is
\beq
\frac{1}{q_j^2 - M_j^2 - i0 \,\eta (q_j-q_i)} \;\;,
\eeq
independently of the value $M_i$ of the mass in the $i$-th line -- the cut line.

In any unitary quantum field theory, the masses of the basic fields are real. If 
some of these fields describe unstable particles, a proper (physical) treatment of 
the corresponding propagators in perturbation theory requires a Dyson summation of 
self-energy insertions,  which produces finite-width effects 
introducing {\em finite} imaginary contributions in the propagators.
A typical form of the ensuing propagator $G_C$ (such as the propagator used in the 
complex-mass scheme\footnote{In the complex-mass scheme, unitarity can be 
	perturbatively recovered (modulo higher-order terms) order by order.} 
\cite{cms}) is 
\beq
\label{comp}
G_C(q;s) = \frac{1}{q^2-s} \;,
\eeq
where $s$ denotes the {\em complex} mass of the unstable particle:
\beq
s = {\rm Re \;} s + i \, {\rm Im \;} s \;\;, 
\quad {\rm with} \;\;\;\; {\rm Re \;} s > 0 > {\rm Im \;} s\; 
\;\;.
\eeq
These complex masses, together with complex couplings, are introduced in both tree-level
and one-loop Feynman diagrams. A natural question that arises in the context of 
the present paper is whether the duality relation between one-loop and phase-space 
integrals (and the FTT, as well) can deal with complex-mass propagators or, more 
generally, with propagators of unstable particles. The answer to this question is 
positive, as we discuss below. 

We consider a one-loop $N$-point scalar integral (see Eq.~(\ref{lng})) where one 
or more of the Feynman propagators of the internal lines are replaced by 
complex-mass propagators $G_C(q_i;s_i)$. To derive a representation of this
one-loop integral in terms of single-cut phase space integrals, we then apply the
same procedure as in Sect.~\ref{sec:dt}. The only difference is 
the presence of the complex-mass 
propagators. In the complex plane of the loop integration variable $q_0$, the 
complex-mass propagators produce poles that are located far off the real axis, 
the displacement being controlled by the finite imaginary part 
of the complex masses. Using the Cauchy theorem as in Eq.~(\ref{ln4}), we derive 
a duality relation that is analogous to Eq.~(\ref{ldt}). The only difference is 
that the the right-hand side of Eq.~(\ref{ldt}) has to be modified:
\beq
\label{repdt}
{\widetilde L}^{(N)}(p_1, p_2, \dots, p_N) \to \;
{\widetilde L}^{(N)}(p_1, p_2, \dots, p_N) +
{\widetilde L}_C^{(N)}(p_1, p_2, \dots, p_N) \;\;.
\eeq
Here, ${\widetilde L}^{(N)}$ denotes the terms that correspond to the residues 
at the poles of the Feynman propagators of the loop integral, while 
${\widetilde L}_C^{(N)}$ denotes those from the poles of the complex-mass 
propagators. 

${\widetilde L}^{(N)}$ is thus expressed as
\beq
\label{dcutf}
{\widetilde L}^{(N)}(p_1, p_2, \dots, p_N) = \int_q \;\; \sum_{i \in F}
\; \td(q_i;M_i) \;
\left[ \;\prod_{{j\neq i}} \;\;\dots
\right]  \;\;,
\eeq
where the sum refers to the internal lines $i$ of the loop with a Feynman
propagator (we use the notation $i \in F$ to denote these cut lines). The term in 
the square bracket denotes the product of the propagators of the uncut lines.
The Feynman propagators of the loop are replaced by the corresponding dual 
propagators (as in Eq.~(\ref{dcut})), while the complex-mass propagators are
{\em unchanged}\footnote{
	The dual propagators arise from the infinitesimal $i0$ displacement 
	produced by the residue at the pole of the Feynman propagator,
	see Sect.~\ref{sec:dt} and Appendix~\ref{app:a}. This infinitesimal 
	imaginary displacement has no effect on the complex-mass propagators, 
	owing to the finite imaginary part of the complex mass.}.  

The expression of ${\widetilde L}_C^{(N)}$ is similar to Eq.~(\ref{dcutf}), but 
the cut lines $i$ are those with complex-mass propagators (we use the notation 
$i \in C$ to denote these cut lines). Taken together
\beeq
{\widetilde L}_C^{(N)}(p_1, p_2, \dots, p_N) &=& \int_q \;\; \sum_{i \in C}
\; \td(q_i;s_i) \;
\left[ \;\prod_{{j\neq i}} \;\;\dots
\right]  \nn \\
\label{dcutc}
&=& \int \frac{d^{d-1}{\bf q}}{(2\pi)^{d-1}} \;\; \sum_{i \in C}
\;\frac{1}{2 {\sqrt {{\bf q}_i^2 +s_i}}}
\;\left[ \;\prod_{{j\neq i}} \;\;\dots 
\right]_{q_{i0}^{}={\sqrt {{\bf q}_i^2 +s_i}}} \;\;,
\eeeq
where the term in the square bracket contains the propagators of the uncut lines. 
Note that in the integral representation on the first line of 
Eq.~(\ref{dcutc}) the `on-shell' delta function $\td(q_i;s_i)$ of the cut
propagator has a formal meaning, since it singles out the residue at the 
complex-mass pole, $q_{i0}^{}=q_{i0}^{(C,+)}= {\sqrt {{\bf q}_i^2 +s_i}}$, which 
has a {\em finite} (and negative) imaginary part. The explicit expression of  
${\widetilde L}_C^{(N)}$ is thus given in the second line of Eq.~(\ref{dcutc}).
Owing to the finite imaginary component of $q_{i0}^{(C,+)}$, 
we can remove the $i0$ prescription in any of the Feynman propagators inside the 
square bracket.

The outcome of our discussion of the duality relation can also be 
used to explain how the FTT can be generalized to deal with complex-mass 
propagators of the internal lines. Following the derivation of the FTT in 
Sect.~\ref{sec:ft}, we can replace the advanced one-loop integral $L_A^{(N)}$ of 
Eq.~(\ref{lna}) with a one-loop integral that contains both advanced propagators 
and complex-mass propagators. This one-loop integral can be rewritten in two 
different ways. First (exploiting Eq.~(\ref{gavsg})), 
it can be expressed, as in the right-hand side of Eq.~(\ref{lnavsln}), in terms 
of a linear combination of the required one-loop integral (i.e.~the integral with 
Feynman and complex-mass propagators) and of multiple-cut phase-space integrals 
$L_{\rm{m-cut}}^{(N)}$. Alternatively, it can be evaluated directly 
by applying the Cauchy theorem as in Eq.~(\ref{lna2}). This direct evaluation 
leads to the computation of the residues at the poles of the complex-mass propagators
(the poles of the advanced propagators do not contribute, since they are placed 
outside the integration contour): the computation gives exactly the contribution 
in Eq.~(\ref{dcutc}). Comparing the expressions obtained in these two ways, we 
conclude that the generalization of the FTT to include complex-mass propagators 
is realized by the following replacement in the right-hand side of Eq.~(\ref{lftt}):
\beq
\label{repft}
L_{\rm{1-cut}}^{(N)}(p_1, p_2, \dots, p_N) \to \;
L_{\rm{1-cut}}^{(N)}(p_1, p_2, \dots, p_N) + 
{\widetilde L}_C^{(N)}(p_1, p_2, \dots, p_N) \;\;.
\eeq
Here, $L_{\rm{1-cut}}^{(N)}$ is the usual contribution (see Eq.~(\ref{1cut})) 
emerging from the single cuts of the sole Feynman propagators of the internal lines
(the complex-mass propagators are not cut), while ${\widetilde L}_C^{(N)}$ is given by 
Eq.~(\ref{dcutc}). Note, in particular, that the complex-mass propagators
do {\em not} produce further $m$-cut contributions $(m \geq 2)$ to the FTT
in addition to the real-mass terms $L_{\rm{m-cut}}^{(N)}$ in Eq.~(\ref{lmcut}).
  
We add a final comment on one-loop integrals with unstable internal particles.
The propagator of an unstable particle can have a form that 
differs from the complex-mass propagator in Eq.~(\ref{comp}). We can introduce, 
for instance, a complex mass, $s(q^2)$, that depends on the momentum $q$ of the 
propagator. We can also include a non-resonant component, in addition to the 
resonant contribution of the complex-mass pole. Independently of its specific 
form, the propagator of the unstable particle produces singularities that are 
located at a {\em finite} imaginary distance from the real axis in the complex 
plane of the loop integration variable $q_0$. Such contributions 
can be included in the duality relation and in the FTT by performing the 
replacements in Eq.~(\ref{repdt}) and in Eq.~(\ref{repft}), respectively. In 
general, the term ${\widetilde L}_C^{(N)}$ has a form that differs from  
Eq.~(\ref{dcutc}) and depends on the actual expression of the propagator and,
in particular, on the singularity structure (poles, branch cuts, $\dots$)
of the propagator in the complex plane.

\section{Gauge theories and gauge poles}
\label{sec:gauge}

The quantization of gauge theories requires the introduction of a gauge-fixing
procedure, which specifies the spin polarization vectors of the 
gauge bosons and the ensuing content of (possible) compensating fictitious 
particles (e.g.~the Faddeev--Popov ghosts in unbroken non-Abelian gauge theories,
or the would-be Goldstone bosons in spontaneously broken gauge theories).

The fictitious particles have their own Feynman propagators, which have to be
taken into account when applying either the FTT or the duality relation. 
This is done in a straightforward manner: if some internal lines in a one-loop
integral correspond to fictitious particles, they have to be cut exactly in the
same way as for physical particles. The multiple-cut phase-space integrals of the 
FTT and the single-cut phase-space integral of the duality relation will include 
the contributions from the cuts of the Feynman propagators of these fictitious 
particles.

The impact of the propagators of the gauge particles is more delicate, since 
they introduce `gauge poles'. This point is discussed below.

The propagator of the (spin 1) gauge boson with momentum $q$ is obtained by 
multiplying the customary Feynman propagator $G(q)$ with the tensor 
$d^{\mu \nu}(q)$, which arises from the sum of the spin polarizations. The general 
form of the polarization tensor is
\beq
\label{polten}
d^{\mu \nu}(q) = - g^{\mu \nu} + (\zeta -1) \; {\ell}^{\mu \nu}(q) \,G_G(q) 
\;\;.
\eeq
The second term on the right-hand side is absent only in the 't~Hooft--Feynman 
gauge $(\zeta=1)$. In any other gauge, this term is present and the tensor 
${\ell}^{\mu \nu}(q)$ propagates longitudinal polarizations, which are proportional 
to $q^\mu$, or $q^\nu$, or $q^\mu q^\nu$. On the one hand, the specific form of 
${\ell}^{\mu \nu}(q)$ is not relevant in the context of the following discussion; 
the only relevant point is that ${\ell}^{\mu \nu}(q)$ has a {\em polynomial} 
dependence on the momentum $q$. On the other hand, the factor $G_G(q)$ (we call 
it `gauge-mode' propagator) has a potentially dangerous, non-polynomial dependence 
on $q$ and, in particular, it produces poles with respect to the momentum variable 
$q$.  

When considering one-loop quantities in gauge theories, we deal with one-loop
integrals containing gauge boson propagators as internal lines of the loop. Therefore,
to derive the FTT or the duality relation, we have to consider the effect 
produced by the gauge polarization tensors. In the {\em 't~Hooft--Feynman gauge}
the effect is {\em harmless}: the polarization tensor is simply $-g^{\mu \nu}$
and factorizes off the loop integration. When applying the Cauchy residue 
theorem as in Sects.~\ref{sec:ft} and \ref{sec:dt} in any other gauge, we have 
to take into account the possible additional contributions that arise from the 
presence of the poles of the gauge-mode propagator $G_G(q)$ (the presence of 
polynomial terms from ${\ell}^{\mu \nu}(q)$ does not interfere with the residue 
theorem).
 
We first discuss the case of spontaneously broken gauge theories. Here, the gauge
boson has a finite mass $M$, and the form of the gauge-mode propagator $G_G(q)$ is
\beq
\label{brok}
G_G(q) = \frac{1}{\zeta (q^2 + \,i0) -M^2} \;\;.
\eeq
Considering the unitary gauge ($\zeta=0$), the gauge-mode propagator does not
depend on $q$ and factorizes off the loop integration in any of the one-loop integrals.
Therefore, the {\em unitary gauge} has only inconsequential implications on the 
use of the FTT and the duality relation for one-loop calculations in gauge theories.
If we instead consider a generic renormalizable gauge (or $R_{\zeta}$ gauge) 
with $\zeta \neq 0$, we see that the gauge-mode propagator introduces a pole
when $q^2 = M^2/\zeta - \,i0$. This is an additional pole with respect to the
physical pole (when  $q^2 = M^2 - \,i0$) from the associated Feynman propagator. 
For the extension of the FTT and the duality relation of 
Sects.~\ref{sec:ft} and \ref{sec:dt} to one-loop computations in the 
$R_{\zeta}$ gauge, one has to properly consider the introduction of additional 
single-cut and multiple-cut contributions from gauge-mode propagators. We will not 
pursue this issue any further in the present paper.

We now discuss the case of unbroken gauge theories, where the gauge boson is
massless. We separately consider two classes of gauges: covariant gauges and
physical gauges. 

In covariant gauges, we have 
\beq
\label{cov}
G_G(q) = \frac{1}{q^2 + \,i0} \;\;.
\eeq
Since the gauge-mode propagator $G_G(q)$ is equal to the Feynman propagator, the 
two propagators together generate a {\em second-order} pole when 
$q^2 = -\,i0$. The extension of the FTT and the duality relation of 
Sects.~\ref{sec:ft} and \ref{sec:dt} to hold for one-loop computations 
in covariant gauges requires a proper treatment of the contributions from 
this type of second-order poles\footnote{Of course, this does not apply to 
	the 't~Hooft--Feynman gauge, where $G_G(q)$ is absent.}. 
This issue is not pursued any further in the present paper.
 
In physical gauges, the typical form of the gauge-mode propagator is
\beq
\label{physg}
G_G(q) = \frac{1}{(n\cdot q)^k} \;\;,  \quad \quad  
k= 1 \;\;{\rm or} \;\;2 \;\;,
\eeq
where $n^\mu$  denotes an auxiliary gauge vector. 
We see that $G_G(q)$ leads to a (first- or second-order) pole when $n\cdot q=0$.
In Coulomb gauge we have $n_\mu = (0, {\bf q})$, where ${\bf q}$ is the space 
component of the gauge boson momentum $q_\mu = (q_0^{}, {\bf q})$. 
In axial $(n\cdot A=0)$ or planar gauges, $n^\mu$ is a fixed external vector
and the pole has to be regularized according to a proper prescription
(the precise position of the pole has to be specified by some imaginary
displacement from the real axis), which we do not specify here, 
since its specific form has no effect on the discussion that follows.

We now consider a generic one-loop integral, whose integrand contains gauge-mode
propagators in addition to Feynman propagators. To derive a duality relation by 
using the residue theorem in the complex plane of the variable $q_0$ (as in 
Sect.~\ref{sec:dt}), we have to take into account the possible contributions from 
the poles of the gauge-mode propagators.

In Coulomb gauge, the pole of $G_G(q)$ is located at ${\bf q}^2=0$. Applying the
residue theorem in the $q_0$ plane at fixed values of ${\bf q}$ (see 
Sect.~\ref{sec:dt} and Appendix~\ref{app:a}), the gauge pole does not contribute. 
We conclude that the gauge-mode propagator remains {\em untouched} in going from 
the one-loop integral to its representation as a single-cut dual integral. Note, 
however, that this conclusion follows from having kept ${\bf q}$ fixed while 
performing the integration over $q_0$. Therefore, the auxiliary future-like 
vector $\eta^\mu$ of the duality relation is necessarily {\em fixed} 
(see Appendix~\ref{app:a}) to be $\eta_\mu= (\eta_0, {\bf 0})$, i.e.~aligned along 
the time direction.

In axial or planar gauges, the pole of $G_G(q)$ is located at 
$nq=n_0 q_0 - n_{d-1} q_{d-1} = 0$. Without loosing generality, we can assume
$n_\mu = (n_0, {\bf 0}_\perp, n_{d-1})$ and apply (see Sect.~\ref{sec:dt}) 
the residue theorem in the complex plane $q_0$ at fixed values of the coordinates 
${\bf q}_\perp$ and $q_{d-1}^\prime=q_{d-1} - q_0 \eta_{d-1}/\eta_0$. Setting 
$\eta_{d-1}/\eta_0 = n_0/n_{d-1}$, we have $nq= -n_{d-1} q_{d-1}^\prime$. Hence, 
$G_G(q)$ does not depend on the integration variable $q_0$. We conclude that the 
gauge-mode propagator, {\em including} the regularization prescription of its 
gauge pole, is {\em untouched} in going from the one-loop integral to its 
representation as a single-cut dual integral. Note, however, that we have set 
$\eta_{d-1}/\eta_0 = n_0/n_{d-1}$. Therefore, since the vector $\eta^\mu$  
specifying the dual prescription is future-like, the above conclusion is valid only 
if the gauge vector  $n^\mu$ is either {\em space-like} or {\em light-like} 
$(n^2 \leq 0)$ and, moreover, the dual vector is fixed to be {\em orthogonal} to 
the gauge vector, $n \cdot \eta=0$. These requirements are not fulfilled if 
$n^\mu$ is time-like\footnote{For example, in the axial gauge $A_0=0$, we 
	have $nq=n_0 q_0$, and the pole of the gauge-mode propagator does not 
	decouple from the integration over $q_0$.}.
The derivation of the duality relation in time-like gauges requires to
properly include contributions from cuts of the gauge-polarization tensors
(these contributions depend on the specific regularization of the gauge poles):
this derivation is beyond the scope of this paper.

Our discussion and conclusions regarding the duality relation in physical gauges
can straightforwardly be used to draw similar conclusions on the validity of the
FTT. The only difference is that in the latter case there is no auxiliary dual 
vector $\eta^\mu$. To be precise, in Coulomb gauge and in space-like or light-like
gauges, the FTT is valid in its customary form, without introducing any 
multiple-cut contributions stemming from the gauge-polarization tensors. In time-like 
gauges, the poles of the gauge-polarization tensors can play a role, and their 
effect has to be taken into account when applying the FTT.

\section{Loop-tree duality at the amplitude level}
\label{sec:dam}

In the final part of Sect.~\ref{sec:ft}, we have discussed how the FTT can be
extended to evaluate not only basic one-loop integrals $L^{(N)}$ but also 
complete one-loop quantities (such as Green's functions and
scattering amplitudes). The same reasoning (see also Sects.~\ref{sec:mass} 
and \ref{sec:gauge}) applies to
the extension of the duality relation to the amplitude level.

The analogue of Eq.~(\ref{aftt}) is the following duality relation:
\beq
\label{adt}
{\cal A}^{({\rm 1-loop})} = - \;
{\cal \widetilde A}^{({\rm 1-loop})}
\;\;,
\eeq
where ${\cal A}^{({\rm 1-loop})}$ generically denotes a one-loop quantity.
The expression ${\cal \widetilde A}^{({\rm 1-loop})}$ on the right-hand side
of Eq.~(\ref{adt}) is obtained in the same way as ${\widetilde L}^{(N)}$
in Eqs.~(\ref{ldt}) and (\ref{dcut}). We start from any Feynman 
diagram in ${\cal A}^{({\rm 1-loop})}$ and consider all possible
replacements of each Feynman propagator $G(q_i)$ of its loop internal lines with
the cut propagator $\td(q_i;M_i)$; the uncut Feynman propagators in the loop
are then replaced by the corresponding dual propagators.
All the other factors in the Feynman diagrams are left unchanged by going from
${\cal A}^{({\rm 1-loop})}$ to ${\cal \widetilde A}^{({\rm 1-loop})}$.

The duality relation (\ref{adt}) is valid in any field theory that is unitary
and local. Some words of caution are, however, needed (see the conclusions
of Sect.~\ref{sec:gauge}) about its applicability to theories with local gauge 
symmetries. In spontaneously broken gauge theories, the duality relation is valid 
in the 't~Hooft--Feynman gauge and in the unitary gauge. In unbroken gauge 
theories, the duality relation is valid in the 't~Hooft--Feynman gauge; it is 
also valid in physical gauges specified by a gauge vector $n^\nu$, provided the 
auxiliary duality vector $\eta^\mu$ is chosen such that  $n\cdot \eta=0$
(this excludes gauges where $n^\nu$ is time-like).

Equation (\ref{adt}) establishes a correspondence between one-loop Feynman
diagrams and the phase-space integral of tree-level Feynman diagrams. The 
right-hand side of Eq.~(\ref{adt}) can be written in the following sketchy form:
\beq
\label{sketch}
{\cal A}^{({\rm 1-loop})} \sim \int_q \;\sum_{P} \;\td(q;M_P) 
\;\sum_{{\rm d.o.f.}(P)} \; {\cal A}^{({\rm tree})}_P \;\;,
\eeq
where $\sum_{P}$ denotes the sum over the particles that can propagate in the 
loop internal lines that are cut, and $\sum_{{\rm d.o.f.}(P)}$ denotes the 
sum over the degrees of freedom (such as spin, colors, ..) of the particle $P$.
The integrand ${\cal A}^{({\rm tree})}_P$ is given by the sum of the tree-level
Feynman diagrams that are obtained by cutting the one-loop Feynman diagrams on 
the left-hand side.

The structure of Eq.~(\ref{sketch}) implies a natural question\footnote{Issues
	related to similar questions were discussed by Feynman \cite{F2} in the 
	context of the FTT.}. 
If ${\cal A}^{({\rm 1-loop})}$ is the one-loop expression of a specific quantity 
${\cal A}$, how is ${\cal A}^{({\rm tree})}_P$ related to the tree-level 
expression ${\cal A}^{({\rm tree})}$ of the same quantity ${\cal A}$? In the next subsections, we show how 
the duality relation can be formulated directly at the amplitude level, when the 
quantity ${\cal A}$ is a Green's function. We also discuss the case of on-shell 
scattering amplitudes.

\subsection{Green's functions}
\label{sec:gfun}

In the following, ${\cal A}_N(p_1, \dots, p_N)$ denotes a generic off-shell 
Green's function with $N$ external lines (the outgoing momentum of the $i$-th line 
is $p_i$). To be precise, we consider Green's functions that are {\em connected} 
and {\em amputated} of the free propagators of the external lines. The tree-level 
and one-loop expressions of ${\cal A}$ are ${\cal A}^{({\rm tree})}$ and 
${\cal A}^{({\rm 1-loop})}$, respectively. The tree-level scattering amplitude 
for a given physical process is obtained from 
${\cal A}^{({\rm tree})}(p_1, \dots, p_N)$ by setting the external momenta on 
their physical mass shell ($p_i^2=M_i^2$, $p_{i 0} \geq 0$ for an outgoing 
particle, $- p_{i 0} \geq 0$ for an incoming particle) and including the 
appropriate wave-function factors of the external particles. The one-loop 
scattering amplitude is obtained from ${\cal A}^{({\rm 1-loop})}$ by specifying 
the renormalization procedure.

To simplify the illustration of the duality relation, we first consider the case
with only one type of massive scalar particles. We thus refer to a theory with a
single real scalar field $\phi$ ($\phi^*=\phi)$ of mass $M$. The particles are
self-interacting through polynomial interactions (e.g. $\phi^3$ or $\phi^4$).
In this case, the duality relation (\ref{sketch}) has the following
explicit form:
\beq
\label{adscalar}
{\cal A}_N^{({\rm 1-loop})}(p_1, \dots, p_N) = + \,\frac{1}{2} 
\;\int \frac{d^d q}{(2\pi)^{d-1}} \;\delta_+(q^2-M^2) \;
{\cal \widetilde A}_{N+2}^{({\rm tree})}(q,-q,p_1, \dots, p_N)
\;\;,
\eeq
where the integrand factor ${\cal A}^{({\rm tree})}$ on the right-hand side
is exactly the tree-level counterpart of the one-loop quantity  
${\cal A}_N^{({\rm 1-loop})}$ on the left-hand side. The tree-level counterpart
${\cal A}_{N+2}^{({\rm tree})}$ involves two additional external lines 
with outgoing momenta $q$ and $-q$.

The tilde superscript in ${\cal \widetilde A}^{({\rm tree})}$ denotes the 
replacement of some of the Feynman propagators with dual propagators. More precisely, to obtain  
${\cal \widetilde A}^{({\rm tree})}(q,-q,\dots)$ from 
${\cal A}^{({\rm tree})}(q,-q, \dots)$, we assign a dual propagator (rather than 
a Feynman propagator) to each internal line with momentum $q+k_j$ ($k_j$ is a linear
combination of the external momenta $p_i$). We note that this step can also be 
performed by using a short-cut recipe, namely  by applying the momentum shift 
$q^\mu \to q^\mu - i0 \,\eta^\mu/(2\eta q)$ in the Feynman propagators of 
${\cal A}^{({\rm tree})}(q,-q, \dots)$.

The momenta $q$ and $-q$ of the two additional external lines of 
${\cal A}_{N+2}^{({\rm tree})}(q,-q,\dots)$ in Eq.~(\ref{adscalar}) are on their 
physical mass-shell: in this respect, ${\cal A}_{N+2}^{({\rm tree})}$ is a 
scattering amplitude (there are no wave-function factors for scalar particles).
More precisely, ${\cal A}_{N+2}^{({\rm tree})}(q,-q, \dots)$ is the 
tree-level physical amplitude that corresponds to the {\em forward-scattering 
process} of a particle with momentum $q$ in the external field produced by $N$ 
self-interacting sources (the $N$ external legs).  

In a theory with different types of particles and antiparticles, the
generalization of Eq.~(\ref{adscalar}) is obtained by including a sum over the
particle types $P$. We find:
\beq
\label{adgen}
{\cal A}_N^{({\rm 1-loop})}(\dots) = + \,\frac{1}{2} 
\;\int \frac{d^d q}{(2\pi)^{d-1}} \;\sum_{P} \;\delta_+(q^2-M^2_P) 
\;\,\sigma(P) \;\,
{\cal \widetilde A}_{N+2}^{({\rm tree})}(P(q) \leftarrow P(q),\dots)
\;\;,
\eeq
where the momenta $p_i$ of $N$ external legs are denoted by `dots', since they 
play no active role on both sides of the equation. Note that $\sum_{P}$ includes the
sum over {\em both} particles and antiparticles (if $P \neq {\overline P}$).
The coefficient $\sigma(P)$ on the right-hand side of Eq.~(\ref{adgen}) is a 
Bose--Fermi statistics factor: $\sigma(P)=+1$ if $P$ is a bosonic particle
(e.g. spin 0 Higgs boson, spin 1 gauge boson), and  $\sigma(P)=-1$ if $P$ is a 
fermionic particle (e.g. spin 1/2 fermion, Faddeev--Popov ghost). 

As in Eq.~(\ref{adscalar}), 
${\cal \widetilde A}^{({\rm tree})}(P(q) \leftarrow P(q),\dots)$ is obtained
from ${\cal A}^{({\rm tree})}(P(q) \leftarrow P(q),\dots)$ by the replacement
of Feynman propagators with dual propagators.
The tree-level expression 
${\cal A}_{N+2}^{({\rm tree})}(P(q) \leftarrow P(q),\dots)$ is the amplitude 
for the forward-scattering process $P(q) \to P(q)$ in the field of the $N$ 
external legs. This expression is obtained from the Green's function
${\cal A}_{N+2}^{({\rm tree})}(P(q), {\bar P}(-q), \dots)$ by setting the
momentum $q$ on the physical mass-shell $(q^2=M^2_P, \;q_0 \geq 0)$ and including
the proper wave-function factors of the external legs with outgoing momenta $q$
and $-q$. We can write:
\beq
\label{forsca}
{\cal A}_{N+2}^{({\rm tree})}(P(q) \leftarrow P(q),\dots) = 
\sum_{{\rm spin, \;color,\,..}} \;\langle P(q)\,| 
\;{\cal A}_{N+2}^{({\rm tree})}(P(q), {\bar P}(-q), \dots) \;|\,P(q)\;\rangle \;,
\eeq
where the (`ket' and `bra') vectors $|\,P(q)\;\rangle$ and  $\langle P(q)\,|$
generically denote the (spin-dependent, color-dependent, ...) 
incoming and outgoing wave-function 
factors of the forward-scattered particle $P$. The quantum numbers 
(spin, color, ...) of the incoming and outgoing wave functions are fixed to be
equal, and the notation $\sum_{{\rm spin, \;color,\,..}}$ denotes the coherent
sum over them.

We illustrate the general notation in Eq.~(\ref{forsca}) with a few explicit 
examples:
\begin{itemize}
\item $P$= gluon ($\lambda$ labels the spin-polarization or helicity states; 
$\mu,\nu$ are Lorentz indices; $a,b$ are color indices) yields 
\beeq
\label{gluon}
\!\!\!\!\!\!
{\cal A}_{N+2}^{({\rm tree})}(g(q) \leftarrow g(q), \dots) \!&=&
\sum_{\lambda} \;\sum_{\mu,\nu} \;\sum_{a,b}
\;\left( {\varepsilon}^{(\lambda)}_{\mu}(q) \right)^*
\,\bigl[ {\cal A}_{N+2}^{({\rm tree})}(g(q), g(-q), 
\dots) \bigr]^{\mu \,\nu}_{a b}
\;{\varepsilon}^{(\lambda)}_{\nu}(q) \nn \\
\!&=& \sum_{\mu,\nu} \;d_{\mu \nu}(q) \;\;\sum_{a,b}
\;\bigl[ {\cal A}_{N+2}^{({\rm tree})}(g(q), g(-q),\dots) 
\bigr]^{\mu \,\nu}_{a b} \;\;,
\eeeq
where ${\varepsilon}^{(\lambda)}_{\nu}(q)$ is the gluon-polarization vector
and $d_{\mu \nu}(q)=\sum_{\lambda} 
( {\varepsilon}^{(\lambda)}_{\mu}(q) )^*
{\varepsilon}^{(\lambda)}_{\nu}(q)$ 
is the corresponding polarization tensor;

\item $P$= massive quark ($s$ labels the spin; $\alpha, \beta$ are Dirac indices; 
$i,j$ are color indices) yields
\beeq
\label{quark}
\!\!\!\!\!\!
{\cal A}_{N+2}^{({\rm tree})}(Q(q) \leftarrow Q(q), \dots) \!&=&
\sum_{s=1,2} \;\sum_{\alpha,\beta} \;\sum_{i,j} \;{\bar u}^{(s)}_{\alpha}(q)
\,\bigl[ {\cal A}_{N+2}^{({\rm tree})}(Q(q), {\bar Q}(-q), \dots) 
\bigr]_{\alpha \,\beta}^{i j}
\;{u}^{(s)}_{\beta}(q) \nn \\
\!&=& \tr \;
\Bigl[
( \slash q + M ) \;
\sum_{i,j} 
\;\bigl[ {\cal A}_{N+2}^{({\rm tree})}(Q(q), {\bar Q}(-q),\dots) \bigr]^{i j}
\Bigr] 
\;\;,
\eeeq
where ${u}^{(s)}_{\beta}(q)$ is the customary Dirac spinor for spin 1/2 fermions;

\item $P$= massive anti-quark ($s$ labels the spin; $\alpha, \beta$ are Dirac 
indices; $i,j$ are color indices) yields
\beeq
\label{antiquark}
\!\!\!\!\!\!\!
{\cal A}_{N+2}^{({\rm tree})}({\bar Q}(q) \leftarrow {\bar Q}(q), \dots) \!\!&=&
- \sum_{s=1,2} \;\sum_{\alpha,\beta} \;\sum_{i,j} \;{\bar v}^{(s)}_{\alpha}(q)
\,\bigl[ {\cal A}_{N+2}^{({\rm tree})}(Q(-q), {\bar Q}(q), \dots) 
\bigr]_{\alpha \,\beta}^{i j}
\;{v}^{(s)}_{\beta}(q) \nn \\
\!\!&=& - \; \tr \;\Bigl[ ( \slash q - M ) \;
\sum_{i,j} 
\;\bigl[ {\cal A}_{N+2}^{({\rm tree})}(Q(-q), {\bar Q}(q), \dots) \bigr]^{i j}
\Bigr]
\;\;, 
\eeeq
where ${v}^{(s)}_{\beta}(q)$ is the customary Dirac spinor for spin 1/2 
anti-fermions.
\end{itemize}

Note that, as stated below Eq.~(\ref{adgen}), 
we sum over both particles and antiparticles.
However, on the right-hand side of Eq.~(\ref{adgen}), $\sum_{P}$ can equivalently 
be defined to just refer to the sum over particles. According to this alternative 
definition, the antiparticle contribution
${\cal \widetilde A}_{N+2}^{({\rm tree})}({\bar P}(q) \leftarrow {\bar P}(q),\dots)$
is absent, and the corresponding particle contribution
${\cal \widetilde A}_{N+2}^{({\rm tree})}({P}(q) \leftarrow {P}(q),\dots)$ must be 
multiplied by a factor of 2.
In view of the issue discussed in
Appendix~\ref{app:c}, the definition of $\sum_{P}$ as sum over both particle 
and antiparticle contributions has to be preferred on general grounds.

We recall that, at the level of one-loop computations, 
the definition of dimensional regularization involves some 
arbitrariness. Although the loop momentum $q^\mu$ is $d$-dimensional,
there is still freedom in the definition of the dimensionality of the momenta of
the external particles and of the number of polarizations of both internal
and external particles. As remarked below Eq.~(\ref{adt}), 
the duality relation acts only on the Feynman propagators of the loop, leaving
unchanged all the other factors in the Feynman diagrams.
Therefore, the dimensional-regularization rules to be used in
the tree-level integrand 
${\cal \widetilde A}_{N+2}^{({\rm tree})}(P(q) \leftarrow P(q),\dots)$ of
Eq.~(\ref{adgen}) are exactly the same as specified in the definition of 
${\cal A}_N^{({\rm 1-loop})}(\dots)$.

We remark that in Eq.~(\ref{adgen}) the on-shell integration momentum $q^\mu$
has always to be considered as $d$-dimensional, with $d$ arbitrary in the 
sense of dimensional regularization. In particular, a $d$-dimensional on-shell
momentum $q^\mu$ is required also if the one-loop Green's function 
${\cal A}_N^{({\rm 1-loop})}$ is finite\footnote{If ${\cal A}_N^{({\rm 1-loop})}$ is finite, the $d$-dimensionality of $q^\mu$ 
in Eq.~(\ref{adgen}) plays simply the role of an intermediate computational 
tool, rather than the role of a necessary regularization procedure. The same
intermediate computational tool is used in other methods to perform
one-loop calculations \cite{Bern:2007dw}: the customary reduction of tensor
integrals to scalar integrals has to be carried out in terms of $d$-dimensional
one-loop integrals; the computation of finite rational terms in one-loop
amplitudes can be carried out by exploiting $d$-dimensional unitarity
techniques.}
(i.e. if it has no infrared and ultraviolet
divergences) in the original and fixed dimensionality (e.g. $d=4$) of the
space-time. The use of a $d$-dimensional $q^\mu$ is necessary since, 
in general, the various terms\footnote{Even if some of these terms correspond to the sum
of the single cuts of a finite loop integral, each single-cut contribution may
not be separately finite. Moreover, possible cancellations of the 
singularities from the various single-cut contributions can be locally
(though, not globally) spoiled by 
the loop-momentum shifts  
(compare Eqs.~(\ref{1cut}) or (\ref{1cutsum}) 
with Eqs.~(\ref{dcut}) or (\ref{d1cutsum}))
that are applied to the separate
single-cut terms. In Eq.~(\ref{adgen}) the momentum shifts are implemented to be
able to identify the different cut momenta of the loop with the common external
momentum $q$ of the tree-level expression 
${\cal \widetilde A}_{N+2}^{({\rm tree})}(P(q) \leftarrow P(q),\dots)$.}
in the integrand on the right-hand side of 
Eq.~(\ref{adgen}) are not separately integrable in a fixed number of
space-time dimensions.

\subsection{Scattering amplitudes}
\label{sec:scamp}

To extend the discussion of Sect.~\ref{sec:gfun} to scattering amplitudes, the 
only relevant point to be examined is the on-shell limit of the corresponding 
Green's functions (the introduction of the wave-function factors of the external 
lines is straightforward).

Considering the off-shell Green's function ${\cal A}_N^{({\rm 1-loop})}$, 
we introduce the following decomposition:
\beq
\label{decom}
{\cal A}_N^{({\rm 1-loop})}= {\cal A}_N^{({\rm 1-loop; \;ex.})}
+ {\cal A}_N^{({\rm 1-loop; \;in.})} \;\;,
\eeq
where ${\cal A}_N^{({\rm 1-loop; \;ex.})}$ is the contribution from one-loop 
insertions on the $N$ external lines, while ${\cal A}_N^{({\rm 1-loop; \;in.})}$ 
is the remaining contribution (i.e.~one-loop insertions on internal lines). In 
explicit form, we have
\beq
\label{gfint}
{\cal A}_N^{({\rm 1-loop; \;ex.})}(p_1, \dots, p_N)=
\sum_{j=1}^N \;{\cal A}_2^{({\rm 1-loop})}(p_j, -p_j) 
\frac{i \;D_j(p_j)}{p_j^2-M_j^2 + i0} 
{\cal A}_N^{({\rm tree})}(p_1, \dots, p_N)
\eeq
where $D_j(p_j)$ is the spin-polarization 
factor\footnote{To be explicit, $D_j(p)$ denotes $d_{\mu \nu}(p)$ (cfr.
Eqs.~(\ref{polten}) and (\ref{gluon})) if the $j$-th line refers to a spin 1
particle, whereas $D_j(p)$ denotes ${\slash p} +M$ (cfr. Eq.~(\ref{quark}))
if the $j$-th line refers to a spin $1/2$ particle.}
of the particle in the internal line with momentum $p_j$.

As is well known, ${\cal A}_N^{({\rm 1-loop; \;ex.})}$ cannot directly be evaluated 
on-shell owing to the kinematical singularity arising from its external-line 
propagators (the propagators with momentum $p_j$ in Eq.~(\ref{gfint})). Thus, to 
calculate the one-loop scattering amplitude, ${\cal A}_N^{({\rm 1-loop; \;ex.})}$ 
has to be first evaluated off-shell, then it has to be renormalized (mass and 
wave-function renormalization), before considering its on-shell limit.

In contrast, the one-loop contribution ${\cal A}_N^{({\rm 1-loop; \;in.})}$ can 
directly be computed in the on-shell limit. In particular, we can write a duality 
relation in the form of Eq.~(\ref{adt}):
\beq
\label{adtin}
{\cal A}_N^{({\rm 1-loop; \;in.})} = - \;
{\cal \widetilde A}_N^{({\rm 1-loop; \;in.})}
\;\;.
\eeq
Here, the integrand of the phase-space integral on the right-hand side contains a 
sum of {\em on-shell} tree-level Feynman diagrams (the $N$ external lines are 
on-shell, and the two additional lines from cutting the loop are also on-shell). 
The algebraic computation of the integrand is thus completely analogous to the 
computation of the (on-shell) tree-level scattering amplitude with $N+2$ external 
legs. Having performed the tree-level computation of the integrand, the result can 
be integrated over the single-particle phase-space to obtain the full one-loop term 
${\cal A}_N^{({\rm 1-loop; \;in.})}$.

We point out that the integrand of the phase-space integral on the right-hand side 
of Eq.~(\ref{adtin}) is {\em not} equal (modulo the replacement of Feynman with 
dual propagators) to the tree-level scattering amplitude with $N+2$ external legs. 
This is because a subset of the diagrams that enter the complete tree-level 
scattering amplitude is not included. This subset has been removed by considering
only ${\cal A}_N^{({\rm 1-loop; \;in.})}$, i.e. by removing 
${\cal A}_N^{({\rm 1-loop; \;ex.})}$ from the complete one-loop expression
${\cal A}_N^{({\rm 1-loop})}$.
 
This `missing' subset of tree-level diagrams can be reinserted in the duality
relation. However, as discussed below, this makes more delicate the on-shell
limit.

We consider the internal-line contribution
${\cal A}_N^{({\rm 1-loop; \;in.})}$ before setting the external lines
on-shell. We can write the following duality relation:
\beeq
\label{adsa}
&& \!\!\!\!\!\!\!\!\!\!\!\!
{\cal A}_N^{({\rm 1-loop; \;in.})}(p_1,\dots,p_N) = + \,\frac{1}{2} 
\;\int \frac{d^d q}{(2\pi)^{d-1}} \;\sum_{P} \;\delta_+(q^2-M^2_P) 
\;\,\sigma(P) \nonumber \\
&&
\times \Bigl\{ \frac{}{}
{\cal \widetilde A}_{N+2}^{({\rm tree})}(P(q) \leftarrow P(q),p_1,\dots,p_N)
\Bigr. \\
&&\Bigl. \quad \;\;\;
- \sum_{j=1}^N 
\;{\cal \widetilde A}_{4}^{({\rm tree})}(P(q) \leftarrow P(q),p_j, -p_j) 
\frac{i \;D_j(p_j)}{p_j^2-M_j^2 + i0} 
{\cal A}_N^{({\rm tree})}(p_1, \dots, p_N)
\Bigr\}
\;\;. \nn
\eeeq
The derivation of this equation is simple.
We first use Eq.~(\ref{decom}) 
to express ${\cal A}_N^{({\rm 1-loop; \;in.})}$ as
difference of ${\cal A}_N^{({\rm 1-loop})}$ and 
${\cal A}_N^{({\rm 1-loop; \;ex.})}$. Then we use Eq.~(\ref{gfint})
to rewrite
${\cal A}_N^{({\rm 1-loop; \;ex.})}$ in terms of ${\cal A}_2^{({\rm 1-loop})}$.
Finally, we express the full one-loop Green's functions 
${\cal A}_N^{({\rm 1-loop})}$ and ${\cal A}_2^{({\rm 1-loop})}$ in terms
of the duality relation (\ref{adgen}).

The duality relation (\ref{adsa}) involves the phase-space integration
of complete tree-level Green's functions, namely 
${\cal A}_N^{({\rm tree})}(p_1, \dots, p_N)$, and 
(the duality-propagator version of) \\
${\cal A}_{N+2}^{({\rm tree})}(P(q) \leftarrow P(q),p_1,\dots,p_N)$
and ${\cal  A}_{4}^{({\rm tree})}(P(q) \leftarrow P(q),p_j, -p_j)$.
The integrand factor in the curly bracket on the right-hand side is well defined 
in the on-shell limit. However, the two terms in the curly bracket
are separately singular in the on-shell limit.
The singularity is purely kinematical; it
simply arises from the propagators of the lines with momenta equal
to the momenta $p_j$ of the external lines. Various procedures
can be devised to introduce an intermediate regularization of the separate
singularities, so as to directly evaluate the two terms close to 
on-shell kinematical configurations.

\section{Final remarks}
\label{sec:fin}

Applying directly the Cauchy residue theorem in the complex plane of any of the 
space-time coordinates of the loop momentum we have derived a duality relation 
between one-loop integrals and single-cut phase-space integrals. The calculation 
of the residues is elementary, but introduces several subtleties. The location 
in the complex plane of the pole of the cut propagator modifies the original 
$+i0$ Feynman prescription of the uncut propagators. One-loop integrals are then 
written as a linear combination of $N$ single-cut phase-space integrals, with 
propagators regularized by a new complex Lorentz-covariant prescription, named 
dual prescription. It is defined through a future-like auxiliary vector $\eta$.
This simple modification compensates for the absence of multiple-cut contributions 
that appear in the FTT. The dependence on $\eta$ cancels, as expected, in the sum 
of all the single-cut contributions, leading to $\eta$-independent results. 

We have generalized the duality relation for internal massive propagators and 
unstable particles. Real masses just modify the position of the poles in the 
complex plane by a translation parallel to the real axis, and thus do not affect 
the dual prescription. Unstable particles introduce a finite imaginary contribution 
in their propagators. The poles of the complex-mass propagators are located at a 
finite imaginary distance from the real axis, and the $+i0$ prescription of the 
usual Feynman propagators can be removed when propagators of unstable particles 
are cut. 

Particular care has to be taken with gauge propagators in both the FTT and the 
duality relation owing to the presence of unphysical extra gauge poles. We have 
discussed this issue, and have identified the different gauge choices where the 
duality relation can be applied in its original form, 
which includes the sole single-cut terms from the Feynman propagators.
This avoids the introduction of additional single-cut terms from the absorptive
contribution of unphysical gauge poles.

Finally, we have extended the duality relation from Feynman integrals to
Green's functions and scattering amplitudes. One-loop scattering amplitudes 
can be obtained starting from tree-level scattering amplitudes
(or, more precisely, from Feynman
diagrams that enter the computation of tree-level scattering amplitudes),
where (some of) the internal propagators are replaced by dual propagators.
This tree-level counterpart is then integrated over a single-particle
phase space to get the one-loop scattering amplitude.

In recent years much progress \cite{Mangano:1990by,off,on} has been achieved 
on the computation
of tree-level amplitudes, including results in compact analytic form. Using the
duality relation, this amount of information at the tree level can be exploited
for applications to analytic calculations at the one-loop level.

The computation of cross sections at next-to-leading order (NLO) requires
the separate evaluation of real and virtual radiative corrections.
Real (virtual) radiative corrections are given by multileg tree-level (one-loop)
matrix elements to be integrated over the multiparticle phase-space of the
physical process.
The loop--tree duality discussed in this paper, as well as other methods that
relates one-loop and phase-space integrals, have an attractive feature
\cite{Soperetal,meth,Kleinschmidt:2007zz,Moretti:2008jj}: 
they recast the virtual radiative corrections in a form that closely
parallels the contribution of the real radiative corrections. This close
correspondence can help to directly combine real and virtual contributions to
NLO cross sections. In particular, using the duality relation,
we can apply \cite{meth} mixed analytical/numerical
techniques to the evaluation of the one-loop virtual contributions. The 
(infrared or ultraviolet) divergent part of the corresponding dual integrals
can be analitycally evaluated in dimensional regularization. The finite 
part of the dual integrals can be computed numerically, together with  
the finite part of the real emission contribution.
Partial results along these lines are presented in Refs.~\cite{meth,tanju}
and further work is in progress.

The extension of the duality relation from one-loop to two-loop Feynman
diagrams is under investigation \cite{inprep}.

\section*{Acknowledgements} 
Financial support by the Ministerio de Educaci\'on y Ciencia (MEC) under
grants FPA2007-60323 and CPAN (CSD2007-00042), by the European Commission
under contracts FLAVIAnet (MRTN-CT-2006-035482), HEPTOOLS (MRTN-CT-2006-035505), 
and MCnet \\
(MRTN-CT-2006-035606), by the INFN-MEC agreement and by BMBF is 
gratefully acknowledged.

Major fractions of the work were completed while three of us (S.C., F.K., G.R.)
were participating in the Workshop 
{\it Advancing Collider Physics: from Twistors to Monte Carlos} at the 
Galileo Galilei Institute (GGI) for Theoretical Physics in Florence: 
we wish to thank the GGI for its hospitality and the INFN for partial support.

S.C. would like to thank Antonio Bassetto, Michael Kr\"amer, Zoltan Nagy and
Dave Soper for discussions. We wish to thank Gudrun Heinrich
and Christian Schwinn for pointing Ref.~\cite{Kleinschmidt:2007zz} out to
us. 

\appendix
\section{Appendix: Derivation of the duality relation}
\label{app:a}

In Sect.~\ref{sec:dt} we have illustrated the derivation of the 
duality relation in Eqs.~(\ref{ldt}) and (\ref{dcut})
by using the residue theorem. The derivation is simple. However,
it involves some subtle points. 
These points are discussed in detail in this Appendix.

Applying the residue theorem in the complex plane of the variable $q_0$, 
the computation of the one-loop integral $L^{(N)}$ reduces to the evaluation of
the residues at $N$ poles, according to Eqs.~(\ref{ln4}) and (\ref{ipole}).

The evaluation of 
the residues in Eq.~(\ref{ipole})
is a key 
point in the derivation 
of the duality relation.
To make this point as clear as possible, we first introduce the notation
$q_{i0}^{(+)}$ to explicitly denote the location of the $i$-th pole,
i.e. the location of the pole with negative imaginary part
(see Eq.~(\ref{fpole})) that is 
produced by the propagator $G(q_i)$.
We 
further simplify our
notation with respect to the explicit dependence on the subscripts that label
the momenta. We write $G(q_j)= G(q_i + (q_j-q_i))$, where
$q_i$ depends on the loop momentum while $(q_j-q_i)=k_{ji}$ is a linear
combination of the external momenta (see Eq.~(\ref{defqi})). Therefore,
to carry out
the explicit computation of the $i$-th residue in Eq.~(\ref{ipole}),
we re-label 
the momenta by $q_i \to q$ and $q_j \to q + k_j$, and we simply evaluate the 
term
\beq
\label{resid}
\left[ {\rm Res}_{\{q_0^{~}=q_{0}^{(+)}\}}
\,G(q) \right]
\;\left[ \;\prod_{j} \,G(q+ k_j) 
\right]_{q_0^{~}=q_{0}^{(+)}} \;\;,
\eeq
where (see Eq.~(\ref{fpole}))
\beq
q_0^{(+)} = {\sqrt {{\bf q}^2 -i0}} \;\;.
\eeq
In the next paragraphs, we follow the steps of Sect.~\ref{sec:dt}
(see Eqs.~(\ref{resGi}) and (\ref{respre})) and we
separately compute the residue of $G(q)$ and its
prefactor -- the associated factor arising from the propagators $G(q+ k_j)$.

The computation of the residue of $G(q)$ gives
\beeq
{\rm Res}_{\{q_0^{~}=q_{0}^{(+)}\}} \,G(q) 
&=& \lim_{q_0^{~} \,\to \,q_{0}^{(+)}} 
\; \left\{ (q_0^{~} - q_{0}^{(+)}) \,\frac{1}{q_0^2 - {\bf q}^2 +i0 } \right\}= 
\frac{1}{2 \,q_{0}^{(+)}} 
\nn \\
\label{resg}
&=& \frac{1}{2 {\sqrt {{\bf q}^2}}} = \int dq_0 \;\delta_+(q^2) \;\;,
\eeeq
thus leading to the result in Eq.~(\ref{resGi}).
Note that the first equality in the second line of Eq.~(\ref{resg}) is 
obtained by removing the $i0$ prescription 
from the previous expression. This is fully justified. The term 
$(q_{0}^{(+)})^{-1}=({\sqrt {{\bf q}^2 -i0}})^{-1}$ becomes singular 
when ${\bf q}^2 \to 0$,
and this corresponds to an end-point singularity in the integration over
${\bf q}$: therefore the $i0$ prescription  has no regularization effect on
such end-point singularity. The second equality in the second line of 
Eq.~(\ref{resg}) simply follows from the definition of the on-shell delta
function $\delta_+(q^2)$.

We now consider the evaluation of the residue prefactor (the
second square-bracket factor in 
Eq.~(\ref{resid})). We first recall that the $i0$ prescription of the 
Feynman propagators has played an important role in the application 
(see Eqs.~(\ref{ln4}) and (\ref{resid}))
of the residue theorem to the computation of the loop integral: 
having 
selected the pole with negative imaginary part, $q_0^{~}=q_{0}^{(+)}$, 
the prescription eventually singled out 
the on-shell mode with positive definite energy, $q_0=|{\bf q}|$ (see Eq.~(\ref{resg})). 
However, we observe that the result in 
Eq.~(\ref{resg}) can be obtained by removing (neglecting) the  
$i0$ prescription either in $q_{0}^{(+)}$ ($q_{0}^{(+)} \to |{\bf q}|$) or
in $G(q)$ ($G(q) \to 1/q^2$):
\beq
{\rm Res}_{\{q_0^{~}=q_{0}^{(+)}\}} \,G(q) = {\rm Res}_{\{q_0^{~}=|{\bf q}|\}} 
\,\frac{1}{q^2} 
= \int dq_0 \;\delta_+(q^2) \;\;.
\eeq
Hence, the $i0$ prescription has no effect on the actual
calculation of the residue of the propagator $G(q)$ in Eq.~(\ref{resid}).
On the basis of this observation, 
we 
might assume that the 
$i0$ prescription also has no effect on the calculation of the residue 
prefactor in Eq.~(\ref{resid}), since the propagators $G(q+ k_j)$ are 
not singular when evaluated at the poles of $G(q)$. 
We 
might thus compute the
residue prefactor by removing the $i0$ prescription; 
under this assumption we 
obtain
\beq
\label{prefactnoi0}
\left[ \;\prod_{j} \,G(q+ k_j) 
\right]_{q_0^{~}=q_{0}^{(+)}} \to \left[ \;\prod_{j} \,\frac{1}{(q+ k_j)^2} 
\right]_{q_0^{~}=|{\bf q}|}\;\;.
\eeq
The expression on the right-hand side of Eq.~(\ref{prefactnoi0}) is 
well-defined, but, when inserted (through Eqs.~(\ref{resid}) and 
(\ref{ipole})) in Eq.~(\ref{ln4}), it leads to
an ill-defined result: the integration over ${\bf q}$ is singular at any
phase-space points where the denominator factors $(q+ k_j)^2$ vanish. 
To recover a well-defined result, we have to reintroduce the $i0$ prescription  
in the residue prefactor.  
We 
might thus maintain the $i0$ prescription in the Feynman propagators
$G(q+ k_j)$ and still keeping $q_0$ at its on-shell value $q_0=|{\bf q}|$;
then we  
obtain
\beq
\label{prefactGi0}
\left[ \;\prod_{j} \,G(q+ k_j) 
\right]_{q_0^{~}=q_{0}^{(+)}} \to \left[ \;\prod_{j} 
\,\frac{1}{(q+ k_j)^2 + i0} 
\right]_{q_0^{~}=|{\bf q}|}\;\;.
\eeq
Inserting (through Eqs.~(\ref{resid}) and 
(\ref{ipole})) Eq.~(\ref{resg}) and the right-hand side of 
Eq.~(\ref{prefactGi0}) 
into Eq.~(\ref{ln4}),
we arrive at a well-defined result for the one-loop integral, since
the singularities
from the propagators $1/(q+ k_j)^2$ are now regularized by the Feynman
$i0$ prescription. However, this result for the one-loop integral
is exactly equal
(see Eqs.~(\ref{1cut}) and (\ref{lftt}))
to the {\rm sole} 1-cut contribution, $L_{\rm{1-cut}}$, of the FTT. The ensuing
contradiction with the FTT can be resolved only if the total contribution
from multiple cuts,
$L_{\rm{2-cut}}+L_{\rm{3-cut}}+~\dots$,
to the FTT vanishes; this is obviously unlikely, and it is actually not true as shown
by the explicit one-loop calculations performed in Sect.~\ref{sec:2p}. 

The discussion of the previous paragraph illustrates that the evaluation
of the one-loop integrals by the direct application of the residue theorem
(as in Eq.~(\ref{ln4})) involves some subtleties. The subtleties mainly concern
the correct treatment of the Feynman $i0$ prescription in the calculation 
of the residue prefactors. A consistent treatment
requires the {\em strict}
computation of the residue prefactor in Eq.~(\ref{resid}): 
the $i0$ prescription in both $G(q+ k_j)$ and $q_{0}^{(+)}$ has to be dealt 
with by considering the imaginary part $i0$ as a {\em finite} 
(thus, for instance, $2i0\neq i0$), though
possibly small, quantity; the limit of infinitesimal values of $i0$ 
has to be taken only at the {\em very} end of the computation, 
thus leading to
the interpretation of the
ensuing $i0$ prescription as mathematical distribution.
Applying this strict procedure, we obtain
\beeq
&&\left[ \;\prod_{j} \,G(q+ k_j) 
\right]_{q_0^{~}=q_{0}^{(+)}} =  
\left[ \;\prod_{j} \,\frac{1}{(q+ k_j)^2 + i0} 
\right]_{q_0^{~}=q_{0}^{(+)}} =
\prod_{j} 
\,\frac{1}{2q_{0}^{(+)}k_{j0} - 2{\bf q}\cdot {\bf k}_j+ k_j^2} \nn \\
\label{prefact}
&& \quad = \prod_{j} 
\,\frac{1}{2 |{\bf q}| k_{j0} - 2{\bf q\cdot k}_j+ k_j^2 - i0 k_{j0}/|{\bf q}|}
 = \left[ \prod_{j} 
\,\frac{1}{ 2qk_j+ k_j^2 - 
i0 k_{j0}/q_0} \right]_{q_0=|{\bf q}|}
\;\;.
\eeeq
The last equality on the first line of Eq.~(\ref{prefact}) simply follows from
setting $q_0^{~}=q_{0}^{(+)}$ in the expression on the square-bracket (note,
in particular, that $q^2=-i0$). The first equality on the second line 
follows from  $2q_{0}^{(+)} \simeq 2|{\bf q}| - i0/|{\bf q}|$ (i.e. 
from expanding $q_{0}^{(+)}$ at small values of $i0$).

The result in Eq.~(\ref{prefact}) for the residue prefactor is well-defined
and leads to a well-defined (i.e. non singular) expression once it is inserted 
in Eq.~(\ref{ln4}). The possible singularities from each of
the propagators $1/(q+k_j)^2$ are regularized by the displacement produced by 
the associated imaginary amount $i0 k_{j0}/q_0$. 
Performing the limit of infinitesimal values of $i0$, only the sign 
of the $i0$ prescription (and not its actual magnitude) is relevant. 
Therefore, since $q_0$ is positive,
in Eq.~(\ref{prefact}) we can
perform the replacement $i0 k_{j0}/q_0 \to i0 \,\eta k_j$, where $\eta^\mu$
is the vector $\eta^\mu=(\eta_0, {\bf 0})$ with $\eta_0 > 0$; we finally obtain
\beeq
\label{resideta}
\left[ \;\prod_{j} \,G(q+ k_j) 
\right]_{q_0^{~}=q_{0}^{(+)}} =  \left[ \prod_{j} 
\,\frac{1}{ (q+ k_j)^2 - 
i0 \,\eta k_{j}}
\right]_{q_0=|{\bf q}|}
\;\;,
\eeeq
which is the result in Eq.~(\ref{respre}) (to be precise,
Eq.~(\ref{respre}) is recovered by reintroducing the original labels of the
momenta of the loop integral according to the replacements $q \to q_i$, $k_j 
\to q_j - q_i$, see the discussion above Eq.~(\ref{resid})).

In the following we explain in more detail the origin of the $\eta$ dependence
in the $i0$ prescription of the dual propagators. The explicit calculation 
performed in this Appendix leads to the introduction of the future-like vector
$\eta^\mu=(\eta_0, {\bf 0})$ (see Eqs.~(\ref{prefact}) and (\ref{resideta})).
As discussed in Sect.~\ref{sec:dt}, different future-like vectors can be
introduced by applying the residue theorem in different systems of coordinates.
To clarify this point, we explicitly show the application of the residue theorem
in light-cone coordinates
(see Eq.~(\ref{pslc})) rather than in space-time
coordinates (as in Eq.~(\ref{ln4})). The one-loop integral can then be evaluated as
follows:
\beeq
\label{lnreslc}
L^{(N)}(p_1, p_2, \dots, p_N) &=& \int_{(q_-, {\bf \qt})} \;\;\; \int dq_+ \;
 \;\; \prod_{i=1}^{N} \,G(q_i) \nn \\
\label{ln4lc}
&=& - \,2 \pi i \; \int_{(q_-, {\bf \qt})} 
 \;\;\sum \; \reslc
 \;\left[ \;\prod_{i=1}^{N} \,G(q_i) \right] \;\;, 
\eeeq 
where we have applied the residue theorem by closing the integration contour 
at $\infty$ in the lower half-plane of the complex variable $q_+$ 
(see Figs.~\ref{fvsa} and \ref{contour}).
We can now compute the residues in Eq.~(\ref{ln4lc}) by closely following  
the analogous computation in Eqs.~(\ref{resid}), (\ref{resg}) and 
(\ref{prefact}).

The analogue of the term in Eq.~(\ref{resid}) is
\beq
\label{residlc}
\left[ {\rm Res}_{\{q_+^{~}=q_{+}^{(+)}\}}
\,G(q) \right]
\;\left[ \;\prod_{j} \,G(q+ k_j) 
\right]_{q_+^{~}=q_{+}^{(+)}} \;\;,
\eeq
where $q_{+}^{(+)}$ denotes the location (in the $q_+$ plane) of the pole 
with negative imaginary part that is 
produced by the propagator $G(q)$. Thus (see Eq.~(\ref{fpole})), we have
\beq
q_+^{(+)} = \frac{{\bf \qt}^2 -i0}{2 q_-} \;\;, \quad {\rm with} \quad q_- > 0
\;\;,
\eeq
where the requirement of negative imaginary part leads to the constraint 
$q_- > 0$. 

The computation of the residue of $G(q)$ gives
\beeq
{\rm Res}_{\{q_+^{~}=q_{+}^{(+)}\}} \,G(q) 
&=& \theta(q_-) \;\lim_{\; q_+^{~} \,\to \,q_{+}^{(+)}} 
\; \left\{ (q_+^{~} - q_{+}^{(+)}) \,\frac{1}{2 q_+ q_- - {\bf \qt}^2 +i0 }
\right\}
\nn \\
\label{resglc}
&=& \theta(q_-) \;\frac{1}{2 q_-} = \int dq_+ \;\,\delta_+(q^2) \;\;.
\eeeq
We see that the residue produces the same factor as in Eq.~(\ref{resg}).

The residue prefactor is evaluated by using the same procedure 
as in Eqs.~(\ref{prefact}) and (\ref{resideta}).
We obtain
\beeq
&&\!\!\!\!\!\!\!\!\!\!\!\!\!\!\!\!\!\! \left[ \;\prod_{j} \,G(q+ k_j) 
\right]_{q_+^{~}=q_{+}^{(+)}} =  
\prod_{j} 
\,\frac{1}{2 q_{+}^{(+)}k_{j-} + 2 q_{-} k_{j+} -
2{\bf \qt}\cdot {\bf \kt}_j+ k_j^2} \nn \\
\label{prefactlc}
&&\!\!\!\!\!\!\!\!  
 = \left[ \prod_{j} 
\,\frac{1}{ 2qk_j+ k_j^2 - 
i0 k_{j-}/q_-} \right]_{q_+={\bf \qt}^2/q_-}
= \left[ \prod_{j} 
\,\frac{1}{ (q+ k_j)^2 - 
i0 \,\eta k_{j}} \right]_{q_+={\bf \qt}^2/q_-} .
\eeeq
The last equality in this equation has been found by performing
the limit of infinitesimal values of $i0$, analogously to Eq.~(\ref{resideta}).
Since $q_-$ is positive, we have thus implemented
the replacement $i0 k_{j-}/q_- \to i0 \,\eta k_j$ where, in the present case,
we have introduced the future-like vector
$\eta^\mu=(\eta_+, {\bf 0_{\perp}}, \eta_-=0)$ with
$\eta_+ =\eta_0 {\sqrt 2} > 0$.

It is important to note that, owing to the on-shell condition
$\delta_+(q^2)$, Eqs.~(\ref{resideta}) and (\ref{prefactlc}) have the same form,
although the corresponding auxiliary vectors $\eta^\mu$ are different:
though $\eta_0 > 0$ in both equations, $\eta$ is time-like ($\eta^2 > 0$)
in Eq.~(\ref{resideta}), whereas it is light-like ($\eta^2 = 0)$
in Eq.~(\ref{prefactlc}).

We also note that the use of the residue theorem in the complex plane $q_0$
at fixed values of $q_-$ and $\bf \qt$ leads to a residue prefactor with
exactly the same light-like vector $\eta^\mu$ as in Eq.~(\ref{prefactlc}).

The main features of the calculation presented in this Appendix are very
general: they are valid in any system of coordinates that can be 
used to apply the residue theorem. 
The residue of $G(q)$ always replaces the Feynman
propagator with the corresponding on-shell propagator $\delta_+(q^2)$
(see Eqs.~(\ref{resGi}), (\ref{resg}) and (\ref{resglc})); 
the residue prefactor generates dual propagators
with an auxiliary vector $\eta$ that depends on the specific system of 
coordinates that has been actually employed
(see Eqs.~(\ref{respre}), (\ref{resideta}) and (\ref{prefactlc})).

We conclude this Appendix by briefly describing the derivation 
(by means of the residue theorem) of the generalized duality relation stated in  
Eq.~(\ref{Lngvscut}). The generalized one-loop integral on the left-hand side
contains both Feynman and advanced propagators. Before applying the residue
theorem, we can specify how the infinitesimal limit `$i0 \to 0$' 
is
performed in the two different types of propagators. We rewrite the advanced
propagator as $G_A(q) = [ \, q^2 - i \rho \,\sg(q_0) \, ]^{-1}$ and, evaluating
the one-loop integral, we perform first the limit  $i0 \to 0$ (at fixed $\rho$)
in the Feynman propagators and then the limit  $i\rho \to 0$ in 
the advanced propagators. We apply the residue theorem by closing the
integration contour at $\infty$  in the lower half-plane of the complex variable 
$q_0$, such that the poles of the advanced propagators do not contribute. Performing the
limit  $i0 \to 0$, the Feynman propagators behave exactly
as in the case of the duality relation
in Eqs.~(\ref{ldt}) and (\ref{dcut}), while the advanced
propagators remain unchanged (since $\rho$ is kept finite). Finally, we perform 
the infinitesimal limit $i\rho \to 0$. We thus obtain Eq.~(\ref{Lngvscut}),
whereas the advanced propagators have not been altered by going from the one-loop
integral on the left-hand side to the phase-space integral on the right-hand
side. 

\setcounter{footnote}{1}
\section{Appendix: An algebraic relation}
\label{app:b}

Here, we provide a proof of the relation (\ref{thetacon}). More generally,
we consider a set of $n$ real variables $\lambda_i$, with $i=1,2,\dots,n$,
that fulfill the constraint
\beq
\label{lacon}
\sum_{i=1}^n \lambda_i =0 \;\;.
\eeq
We shall prove the following relation:
\beq
\theta(\lambda_1) \,\theta(\lambda_1+\lambda_2) \,\dots 
\,\theta(\lambda_1+\lambda_2+ \dots+\lambda_{n-1} ) + {\rm cyclic \;\; perms.}
 = 1 \;\;.
\label{eq:algebraic}
\eeq

Equation (\ref{thetacon}) simply follows from setting $\lambda_i=\eta \,p_i$
and is just a consequence of momentum conservation, namely Eq.~(\ref{lacon}).
Note that the future-like nature of the vector $\eta$ plays no role in 
Eq.~(\ref{thetacon}). 

To present the proof of Eq.~(\ref{eq:algebraic}),  
we first define the following function $F_n$:
\beq
F_n (\lambda_1,\cdots, \lambda_n)
= \theta(\lambda_1) \,\theta(\lambda_1+\lambda_2) \,\dots 
\,\theta(\lambda_1+\lambda_2+ \dots+\lambda_{n-1} ) + {\rm cyclic \;\; perms.}
\;.
\label{eq:cyclic1}
\eeq
Then, we proceed by induction. Assuming that Eq.~(\ref{eq:algebraic}) is valid for 
$n-1$ real variables (i.e. $F_{n-1}=1$),
we shall prove that it is valid for $n$ variables (i.e. $F_{n}=1$).

The proof is simple. We first note two properties: 
owing to Eq.~(\ref{lacon}), 
at least one of the variables $\lambda_i$ must have a positive value;
$F_n (\lambda_1,\cdots, \lambda_n)$
has a fully symmetric dependence on the $n$ variables $\lambda_i$.
If we can show that $F_{n}=1$ when one of the variables, say $\lambda_1$,
is positive, from these two properties it follows that $F_{n}$ is always equal 
to unity.

We consider the various terms on the right-hand side of Eq.~(\ref{eq:cyclic1})
and, setting $\lambda_1 > 0$, we have:
\beq
\label{la1}
\theta(\lambda_1) \,\theta(\lambda_1+\lambda_2) \,\dots 
\,\theta(\lambda_1+\lambda_2+\dots+\lambda_{n-1})
= \theta(\lambda_1+\lambda_2) \,\dots 
\,\theta(\lambda_1+\lambda_2+\dots+\lambda_{n-1}) \;,
\eeq
\beq
\label{la2}
\theta(\lambda_2) \,\theta(\lambda_2+\lambda_3) \;\dots 
\,\theta(\lambda_2+\dots+\lambda_{n}) = 0  \;,
\eeq
\beeq
\label{lai}
&&\theta(\lambda_i) \,\dots \,\theta(\lambda_i+\dots+\lambda_n) 
\,\theta(\lambda_i+\dots+\lambda_n+\lambda_1)
\,\theta(\lambda_i+\dots+\lambda_n+\lambda_1+\lambda_2) 
\,\dots \nn \\
&&= \theta(\lambda_i) \,\dots \,\theta(\lambda_i+\dots+\lambda_n) 
\,\theta(\lambda_i+\dots+\lambda_n+\lambda_1+\lambda_2) 
\,\dots \;\;, \quad (i\geq 3) \;\;.
\eeeq
The equality in Eq.~(\ref{la1}) simply follows from $\theta(\lambda_1)=1$.
To obtain Eq.~(\ref{la2}), we exploit momentum conservation to get
$\theta(\lambda_2+\dots+\lambda_{n})= \theta(-\lambda_1)$, and then we use
$\theta(-\lambda_1)=0$. To obtain Eq.~(\ref{lai}) we simply use
$\theta(\lambda_i+\dots+\lambda_n+\lambda_1)=1$, which follows from the presence
of $\theta(\lambda_i+\dots+\lambda_n)$ and from $\lambda_1 > 0$.

Summing the terms on the left-hand side of Eqs.~(\ref{la1}), (\ref{la2})
and (\ref{lai}), 
we obtain $F_n (\lambda_1, \lambda_2,\cdots, \lambda_n)$;
the sum of the corresponding terms on the right-hand side gives 
$F_{n-1} (\lambda_1+\lambda_2,\cdots, \lambda_n)$ (note that the two variables
$\lambda_1$ and $\lambda_2$ are replaced by the single variable 
$\lambda_1+\lambda_2$). Therefore, 
we obtain~\footnote{Note that, starting from $\lambda_i >0$, we would have obtained \\
$F_n(\cdots, \lambda_i, \lambda_{i+1},\cdots)=
F_{n-1}(\cdots,\lambda_i+\lambda_{i+1},\cdots)$. 
Starting from $\lambda_i < 0$, we can analogously obtain 
$F_n(\cdots, \lambda_{i-1}, \lambda_{i},\cdots)=
F_{n-1}(\cdots,\lambda_{i-1}+\lambda_{i},\cdots)$.}
$F_n (\lambda_1, \lambda_2,\cdots, \lambda_n)=
F_{n-1} (\lambda_1+\lambda_2,\cdots, \lambda_n)$, and hence 
$F_n (\lambda_1, \lambda_2,\cdots, \lambda_n)=1$ from the induction assumption.
This completes the proof of Eq.~(\ref{eq:algebraic}).


\section{Appendix: Tadpoles and off-forward regularization}
\label{app:c}

The one-loop Feynman diagrams that contribute to a generic quantity
include diagrams with tadpoles. Among them
there are also `1-particle 
tadpoles', namely tadpoles
linked to
a single line of the diagram (Fig.~\ref{tad}--{\em left}). This 
single line necessarily corresponds to the {\em zero-momentum}
propagation of a particle $K$
with no associated antiparticle (i.e. the particle $K$ is the quantum of
a real bosonic field). If the particle $K$ is {\em massless}, its zero-momentum
propagator is ill-defined (it gives $1/(+i0)$). In this case,
the theory is consistent (perturbatively stable) only if the 
1-particle tadpole vanishes. 

\begin{figure}[htb]
\begin{center}
\vspace*{18mm}
\begin{picture}(350,110)(0,-10)
\SetWidth{1.2}
%
\GCirc(50,50){30}{0}
\BCirc(50,115){16}
\ArrowArc(50,115)(22,60,190)
\ArrowLine(69.28,72.98)(88.57,95.96)
\ArrowLine(30.71,72.98)(11.43,95.96)
\ArrowLine(75.98,35)(101.96,20)
\ArrowLine(24.02,35)(-1.96,20)
\Line(50,80)(50,100)
\color{blue}
\Text(100,10)[]{$p_{i}$}
\Text(40,144)[]{$q$}
\color{red}
\Text(55,90)[]{$K$}
\Vertex(50,6){1.4}
\Vertex(65.39,11.71){1.4}
\Vertex(32.6,5.71){1.4}%
\SetOffset(210,0)
\GCirc(50,50){30}{0}
\ArrowLine(69.28,72.98)(88.57,95.96)
\ArrowLine(30.71,72.98)(11.43,95.96)
\ArrowLine(75.98,35)(101.96,20)
\ArrowLine(24.02,35)(-1.96,20)
\Line(50,80)(50,100)
\ArrowLine(50,100)(70,120)
\ArrowLine(30,120)(50,100)
\color{blue}
\Text(100,10)[]{$p_{i}$}
\Text(30,130)[]{$q$}
\Text(70,130)[]{$q$}
\color{red}
\Text(55,90)[]{$K$}
\Vertex(50,6){1.4}
\Vertex(65.39,11.71){1.4}
\Vertex(32.6,5.71){1.4}
%
\end{picture}
\end{center}
\caption{\label{tad}
 {\em A one-loop Feynman diagram with a 1-particle tadpole (left), and the
 tree-level diagram that is obtained by cutting the tadpole (right). The black
 disk denotes a generic tree diagram.}}
\end{figure}
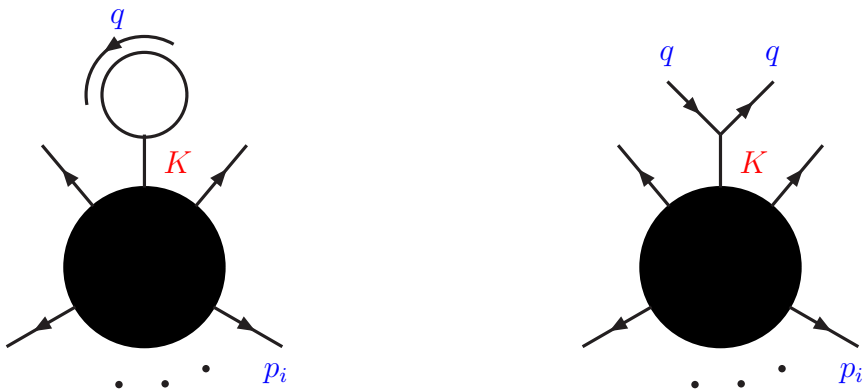

In any consistent theories, the diagrams with 
1-particle tadpoles linked to
a massless line are considered to be vanishing, by definition. Therefore,
they are harmless in any direct computations at one-loop level:
they are simply removed from the set of one-loop diagrams to be computed.
However, their effect may appear to be `dangerous' 
in the context of loop-tree
duality at the amplitude level.

To illustrate the origin of the possible `danger', 
we consider the right-hand
side of the duality
relation in Eq.~(\ref{adgen}). Here, the integrand is related to the tree-level
forward-scattering amplitude 
${\cal A}_{N+2}^{({\rm tree})}(P(q) \leftarrow P(q),\dots)$. This
amplitude is the full tree-level amplutude and, therefore, it includes also 
the tree-level diagrams that are obtained by cutting 
1-particle tadpoles (see Fig.~\ref{tad}--{\em right}). If the 
1-particle tadpole is linked to the ill-defined propagator $1/(+i0)$
of a massless particle $K$, 
the corresponding diagram in the tree-level scattering amplitude
is also ill-defined. To make Eq.~(\ref{adgen}) a well-defined relation,
${\cal \widetilde A}_{N+2}^{({\rm tree})}(P(q) \leftarrow P(q),\dots)$
has to be defined starting from a regularized version of the 
(possibly ill-defined)
amplitude ${\cal A}_{N+2}^{({\rm tree})}(P(q) \leftarrow P(q),\dots)$.
This regularization procedure has to be consistent: the {\em only} effect 
that it can eventually produce in the right-hand of Eq.~(\ref{adgen})
is the cancellation of the terms that correspond to vanishing tadpole diagrams
at one-loop level. 

We introduce a very simple regularization procedure of tadpole-induced 
(forward-scattering) singularities: the two momenta of the on-shell particle
$P$ are displaced slightly off-forward. We thus consider 
the following off-forward scattering amplitude (cf.~Eq.~(\ref{forsca})):
\beq
\label{offforsca}
{\cal A}_{N+2}^{({\rm tree})}(P(q) \leftarrow P(q_1),\dots) = 
\sum_{{\rm spin, \;color,\,..}} \;\langle P(q)\,| 
\;{\cal A}_{N+2}^{({\rm tree})}(P(q), {\bar P}(-q_1), \dots) 
\;|\,P(q_1)\;\rangle \;,
\eeq
where $q \neq q_1$, although both $q$ and $q_1$ are on-shell.
It is important to note that the expression in Eq.~(\ref{offforsca})
includes the wave-function factors of the on-shell external lines 
with momenta $q$ and
$q_1$; in particular, it includes the coherent
sum over the spins and colours of the wave functions of the incoming and
outgoing particles $P$.
The possibly ill-defined propagators $1/(+i0)$, related to 
forward-scattering kinematics, are obviously replaced by $1/((q-q_1)^2+i0)$
when considering 
${\cal A}_{N+2}^{({\rm tree})}(P(q) \leftarrow P(q_1),\dots)$.

As discussed in Sect.~\ref{sec:dam}, the amplitude
${\cal \widetilde A}_{N+2}^{({\rm tree})}(P(q) \leftarrow P(q),\dots)$
is obtained by starting from 
${\cal A}_{N+2}^{({\rm tree})}(P(q) \leftarrow P(q),\dots)$ and replacing
Feynman propagators with dual propagators.
The {\em off-forward regularization} is obtained by starting from the 
corresponding regularized version of 
${\cal A}_{N+2}^{({\rm tree})}(P(q) \leftarrow P(q),\dots)$. The regularized
version is defined as follows:
\vspace*{-3mm}
\begin{itemize}
\item
if $P$ has no corresponding antiparticle, we consider the limit $q_1 \to q$
of \\
${\cal A}_{N+2}^{({\rm tree})}(P(q) \leftarrow P(q_1),\dots)$;
\item
if $P$ has a corresponding antiparticle ${\overline P}$, we first combine the
particle and antiparticle contributions and then we consider the limit 
$q_1 \to q$ of the sum \\
${\cal A}_{N+2}^{({\rm tree})}(P(q) \leftarrow P(q_1),\dots)
+ {\cal A}_{N+2}^{({\rm tree})}({\overline P}(q) 
\leftarrow {\overline P}(q_1),\dots)$.
\end{itemize}
\vspace*{-3mm}
The key point of the off-forward regularization is simple:
rather than considering the forward-scattering limit at fixed values of 
the spin and colour, we first sum over spins, colours and, possibly, particle
and antiparticle, and then we consider the forward-scattering limit.

Within the Standard Model of strong and electroweak interactions,
the massless particles $K$ that can produce tadpole-induced singularities
are gluons and photons (Fig.~\ref{offref}).
We consider these explicit examples to illustrate how the
off-forward regularization consistently leads to the cancellation
of tadpole-induced singularities.

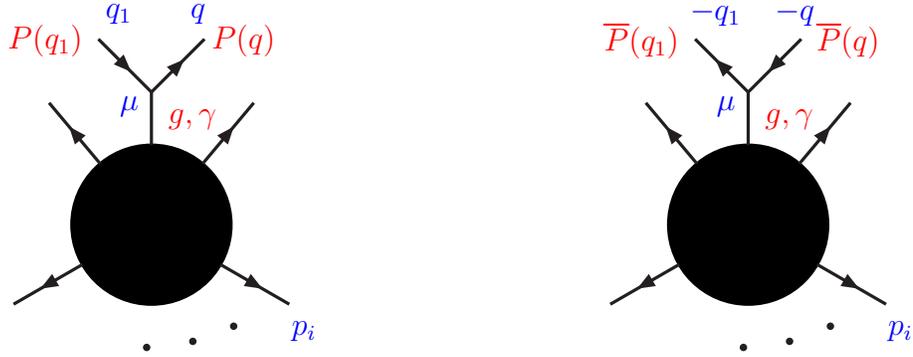
\begin{figure}[htb]
\begin{center}
\vspace*{18mm}
\begin{picture}(350,110)(0,-10)
\SetWidth{1.2}
%
\GCirc(50,50){30}{0}
\ArrowLine(69.28,72.98)(88.57,95.96)
\ArrowLine(30.71,72.98)(11.43,95.96)
\ArrowLine(75.98,35)(101.96,20)
\ArrowLine(24.02,35)(-1.96,20)
\Line(50,80)(50,100)
\ArrowLine(50,100)(70,120)
\ArrowLine(30,120)(50,100)
\color{red}
\Text(10,120)[]{$P(q_1)$}
\Text(85,120)[]{$P(q)$}
\color{blue}
\Text(100,10)[]{$p_{i}$}
\Text(30,130)[]{$q_1$}
\Text(60,130)[]{$q$}
\Text(34,95)[]{$\mu$}
\color{red}
\Text(50,90)[]{$g,\gamma$}
\Vertex(50,6){1.4}
\Vertex(65.39,11.71){1.4}
\Vertex(32.6,3.71){1.4}
%
\SetOffset(210,0)
\GCirc(50,50){30}{0}
\ArrowLine(69.28,72.98)(88.57,95.96)
\ArrowLine(30.71,72.98)(11.43,95.96)
\ArrowLine(75.98,35)(101.96,20)
\ArrowLine(24.02,35)(-1.96,20)
\Line(50,80)(50,100)
\ArrowLine(70,120)(50,100)
\ArrowLine(50,100)(30,120)
\color{red}
\Text(10,120)[]{${\overline P}(q_1)$}
\Text(88,120)[]{${\overline P}(q)$}
\color{blue}
\Text(100,10)[]{$p_{i}$}
\Text(30,130)[]{$- q_1$}
\Text(60,130)[]{$- q$}
\Text(34,95)[]{$\mu$}
\color{red}
\Text(50,90)[]{$g,\gamma$}
\Vertex(50,6){1.4}
\Vertex(65.39,11.71){1.4}
\Vertex(32.6,3.71){1.4}
\end{picture}
\end{center}
\caption{\label{offref}
{\em Off-forward regularization of tree-level diagrams with tadpole-induced
singularities: contributions from particle (left) and antiparticle (right) 
scattering.}}
\end{figure}

The gluon case is very trivial, since the colour sum on the right-hand side of
Eq.~(\ref{offforsca}) directly cancels any tadpole-induced singularities.
The cancellation is eventually the consequence of colour conservation. 
To be precise, the coupling $P(q)P(q_1)g^*$
(see Fig.~\ref{offref}) is proportional
to the colour matrix $T^a_{c c_1}$, where $a$ is the color index of the gluon,
and $c$ and $c_1$ are the colour indeces of $P(q)$ and $P(q_1)$, respectively.
The sum over the colours of the particle $P$ thus gives $\tr (T^a)=0$,
independently of the specific case (gluon, quark, ghost, ..) of particle $P$.

In the photon case, the particle $P$ is charged and thus $P\neq {\overline P}$.
In this case, the cancellation of the tadpole-induced singularity is 
eventually
due to charge conservation, and it is achieved
by summing the contributions of $P$ (Fig.~\ref{offref}--{\em left})
and ${\overline P}$ (Fig.~\ref{offref}--{\em right}). To be precise,
we can consider explicitly the three cases: $P$ is a charged scalar, $P$
is a charged vector boson and $P$ is a charged fermion.

If $P$ is a charged scalar particle, the couplings $P(q)P(q_1)\gamma^*$ 
and ${\overline P}(q){\overline P}(q_1)\gamma^*$
lead
to the factors $(q+q_1)^\mu$ and $-(q+q_1)^\mu$, respectively 
($\mu$ is the Lorentz index of the photon). These two factors simply differ
by the overall sign, and thus they cancel each other.  

If $P$ is a charged vector boson, the cancellation occurs as in the case of
scalar particles. To be precise, the scalar vertex $(q+q_1)^\mu$ is  
replaced by the vertex $\Gamma^{\nu \mu \nu_1}(q,q_1-q,-q_1)=(q+q_1)^\mu
g^{\nu \nu_1}+\dots$, where $\nu$ and $\nu_1$ are the Lorentz indeces of 
the vector bosons $P(q)$ and $P(q_1)$, respectively. Including the
wave-function polarization vectors of the charged vector bosons, we can define
\beq
V^{(\lambda) \mu}(q,q_1) \equiv \sum_{\nu, \nu_1}
({\varepsilon}^{(\lambda)}_{\nu}(q))^* \;\Gamma^{\nu \mu \nu_1}(q,q_1-q,-q_1)
\;{\varepsilon}^{(\lambda)}_{\nu_1}(q_1) \;.
\eeq
The couplings $P(q)P(q_1)\gamma^*$ 
and ${\overline P}(q){\overline P}(q_1)\gamma^*$
lead to the factors $V^{(\lambda) \mu}(q,q_1)$ and 
$- V^{(\lambda) \mu}(q,q_1)$, respectively. Therefore,
these two contributions cancel each other for any {\em fixed}
polarization state
$\lambda$ of the vector boson.

If $P$ is a charged (massive or massless) fermion, 
the cancellation takes place after summing
over the spin states $s=1,2$ of the fermion and antifermion contributions.
Indeed, the sum of the couplings $P(q)P(q_1)\gamma^*$ and 
${\overline P}(q){\overline P}(q_1)\gamma^*$
produces the factor
\beq
\label{tadfer}
\sum_{s=1,2} \;{\bar u}^{(s)}(q) \;\gamma^{\mu} \;{u}^{(s)}(q_1) -
\sum_{s=1,2} \;{\bar v}^{(s)}(q_1) \;\gamma^{\mu} \;{v}^{(s)}(q) \;\;,
\eeq
which identically vanishes. 

To show that the expression in Eq.~(\ref{tadfer})
vanishes, we use the following relations:
\beeq
\label{uubar}
\sum_{s=1,2} \;{u}^{(s)}_{\alpha}(q_1) \;{\bar u}^{(s)}_{\beta}(q) &=&
\left[ \frac{(\slash q_1 + M ) (1+ \gamma_0) 
(\slash q + M )}{2 \sqrt{(q_{10} + M)(q_{0} + M)}} 
\right]_{\alpha \beta} \;, \nn \\
- \sum_{s=1,2} \;{v}^{(s)}_{\alpha}(q) \;{\bar v}^{(s)}_{\beta}(q_1) &=&
\left[ \frac{(- \slash q + M ) (1- \gamma_0) 
(- \slash q_1 + M )}{2 \sqrt{(q_{10} + M)(q_{0} + M)}} 
\right]_{\alpha \beta} \;,
\eeeq

\beeq
\label{ubargu}
\sum_{s=1,2} \;{\bar u}^{(s)}(q) \;\gamma^{\mu} 
\;{u}^{(s)}(q_1)  &=& \frac{ \tr \;\left[ \gamma^{\mu}
(\slash q_1 + M ) (1+ \gamma_0) 
(\slash q + M )\right]}{2 \sqrt{(q_{10} + M)(q_{0} + M)}}
\;, \nn \\
- \sum_{s=1,2} \;{\bar v}^{(s)}(q_1)
\;\gamma^{\mu} \;{v}^{(s)}(q)  &=& \frac{ \tr \;\left[ \gamma^{\mu}
(\slash q - M ) (1- \gamma_0) 
(\slash q_1 - M )\right]}{2 \sqrt{(q_{10} + M)(q_{0} + M)}}
\;,
\eeeq

\beeq
\label{trace}
\tr \;\left[ \gamma^{\mu}
(\slash q_1 + M ) (1+ \gamma_0) 
(\slash q + M )\right] &=&\!\! - \;\tr \;\left[ \gamma^{\mu}
(\slash q - M ) (1- \gamma_0) (\slash q_1 - M )\right] \\
&=&\!\! 4 \left[ M (q_1+q)^\mu + ( q_{0} q_1^\mu +q_{10} q^\mu ) + 
\frac{1}{2} g^{\mu 0} (q-q_1)^2 \nn
\right] \;\;.
\eeeq
The two relations in Eq.~(\ref{uubar}) are directly derived by using the 
explicit expressions of the Dirac spinors ${u}^{(s)}$ and ${v}^{(s)}$ from the
solutions of the Dirac equation. The two relations in Eq.~(\ref{ubargu})
are obtained from Eq.~(\ref{uubar}), and Eq.~(\ref{trace}) is the result of an
elementary computation of Dirac $\gamma$ matrices.
Using the relations in Eqs.~(\ref{ubargu}) and (\ref{trace}), we immediately see
that the expression in Eq.~(\ref{tadfer}) is equal to zero.

\end{document}